\pdfoutput=0
\documentclass[11pt]{article}
\usepackage{axodraw}
\usepackage{epsfig}
\usepackage{amsfonts}
\usepackage{amsmath}
\usepackage{bbm,bm}
\usepackage{cite}
\allowdisplaybreaks[1]

\input pix.sty

\renewcommand{\vec}[1]{{\bf #1}}

\newcommand{\tinymsbar}{{\overline{\mbox{\tiny\rm{MS}}}}}
\newcommand{\Lambdamsbar}{{\Lambda_\tinymsbar}}
\newcommand{\alphas}{\alpha_{\rm s}}
\newcommand{\Nf}{N_{\rm f}}
\newcommand{\Nc}{N_{\rm c}}

\newcommand{\M}{\rmii{$M$}}
\newcommand{\E}{\rmii{$E$}}
\renewcommand{\B}{\rmii{$B$}}
\newcommand{\I}{\mathcal{I}}
\newcommand{\CF}{C_\rmii{F}}

\newcommand{\mE}{m_\rmii{E}}

\newcommand{\mG}{m_\rmiii{G}}

\newcommand{\gammaE}{\gamma_\rmii{E}}

\newcommand{\rmO}{{\mathcal{O}}}
\newcommand{\bmu}{\bar\mu}

\def\lsi{\raise0.3ex\hbox{$<$\kern-0.75em\raise-1.1ex\hbox{$\sim$}}}
\def\gsi{\raise0.3ex\hbox{$>$\kern-0.75em\raise-1.1ex\hbox{$\sim$}}}

\newcommand{\gsim}{\mathop{\gsi}}

\newcommand{\nF}{n_\rmii{F}}
\newcommand{\nB}{n_\rmii{B}}
\newcommand{\rmii}[1]{{\mbox{\tiny\rm{#1}}}}
\newcommand{\rmiii}[1]{{\mbox{\tiny{$\scriptstyle{\rm#1}$}}}}
\newcommand{\re}{\mathop{\mbox{Re}}}
\newcommand{\im}{\mathop{\mbox{Im}}}
\newcommand{\Tint}[1]{{\hbox{$\sum$}\!\!\!\!\!\!\!\int\,}_{\!\!\!\!\raise-0.9ex\hbox{$\scriptstyle{#1}$}}}
\newcommand{\Tinti}[1]{{{\Sigma}\!\!\!\!\raise0.3ex\hbox{$\int$}_\rmii{${#1}$}}}
\newcommand{\Tintip}[1]{{{\Sigma'}\!\!\!\!\!\raise0.3ex\hbox{$\int$}_\rmii{${#1}$}}}

\newcommand{\bi}{\begin{itemize}}
\newcommand{\ei}{\end{itemize}}
\newcommand{\hide}[1]{ }

\def\TAsc(#1,#2)(#3,#4,#5)%
{\SetWidth{2.0}\CArc(#1,#2)(#3,#4,#5)\SetWidth{1.0}}
\def\Lwidth{3}

\def\TAgl(#1,#2)(#3,#4,#5){\SetWidth{2.0}\PhotonArc(#1,#2)(#3,#4,#5){\Lwidth}%
{6.283 #3 mul 360 div #4 #5 sub #4 #5 sub mul sqrt mul Tdensity mul}%
\SetWidth{1.0}}
\def\TLgl(#1,#2)(#3,#4){\SetWidth{2.0}\Photon(#1,#2)(#3,#4){\Lwidth}
{#1 #3 sub #1 #3 sub mul #2 #4 sub #2 #4 sub mul add sqrt Tdensity mul}%
\SetWidth{1.0}}
\def\Lwidth{1.3}

\newcommand{\picu}[1]{\;\parbox[c]{30pt}{\begin{picture}(30,30)(0,0)
\SetWidth{1.0}\SetScale{1.0} #1 \end{picture}}\; }
\def\EleA{\picu{%
 \CArc(15,15)(15,0,360)%
 \Lgl(15,0)(15,30)%
 \COval(15,0)(2,2)(0){Black}{Black}%
 \COval(15,30)(2,2)(0){Black}{Black}%
}}
\def\EleB{\picu{%
 \CArc(15,15)(15,0,360)%
 \Lgl(15,0)(15,30)%
 \COval(15,0)(2,2)(0){Black}{Black}%
 \COval(15,30)(2,2)(0){Black}{Black}%
 \Agl(29,15)(8,100,260)%
}}
\def\EleBB{\picu{%
 \CArc(15,15)(15,0,360)%
 \Lgl(15,0)(15,30)%
 \Lgl(0,15)(12,15)%
 \Lgl(18,15)(30,15)%
 \COval(15,0)(2,2)(0){Black}{Black}%
 \COval(15,30)(2,2)(0){Black}{Black}%
}}
\def\EleC{\picu{%
 \CArc(15,15)(15,0,360)%
 \Agl(7,4)(8,-30,130)%
 \Agl(23,26)(8,150,310)%
 \COval(15,0)(2,2)(0){Black}{Black}%
 \COval(15,30)(2,2)(0){Black}{Black}%
}}
\def\EleD{\picu{%
 \CArc(15,15)(15,0,360)%
 \Agl(23,4)(8,50,210)%
 \Agl(23,26)(8,150,310)%
 \COval(15,0)(2,2)(0){Black}{Black}%
 \COval(15,30)(2,2)(0){Black}{Black}%
}}
\def\EleE{\picu{%
 \CArc(15,15)(15,0,360)%
 \Agl(40,30)(25,180,242)%
 \Agl(40,0)(25,118,137)%
 \Agl(40,0)(25,150,180)%
 \COval(15,0)(2,2)(0){Black}{Black}%
 \COval(15,30)(2,2)(0){Black}{Black}%
}}
\def\EleF{\picu{%
 \CArc(15,15)(15,0,360)%
 \Lgl(15,0)(15,30)%
 \Agl(40,40)(26,205,245)%
 \COval(15,0)(2,2)(0){Black}{Black}%
 \COval(15,30)(2,2)(0){Black}{Black}%
}}
\def\EleG{\picu{%
 \CArc(15,15)(15,0,360)%
 \Agl(40,15)(30,150,210)%
 \Agl(-10,15)(30,-30,30)%
 \COval(15,0)(2,2)(0){Black}{Black}%
 \COval(15,30)(2,2)(0){Black}{Black}%
}}
\def\EleH{\picu{%
 \CArc(15,15)(15,0,360)%
 \Lgl(15,0)(15,12)%
 \Agl(10,21.5)(10,-65,65)%
 \Agl(20,21.5)(10,115,245)%
 \COval(15,0)(2,2)(0){Black}{Black}%
 \COval(15,30)(2,2)(0){Black}{Black}%
}}
\def\EleI{\picu{%
 \CArc(15,15)(15,0,360)%
 \Lgl(15,0)(15,30)%
 \Lgl(15,15)(30,15)%
 \COval(15,0)(2,2)(0){Black}{Black}%
 \COval(15,30)(2,2)(0){Black}{Black}%
}}
\def\EleJ{\picu{%
 \CArc(15,15)(15,0,360)%
 \Lgl(15,0)(15,30)%
 \COval(15,0)(2,2)(0){Black}{Black}%
 \COval(15,30)(2,2)(0){Black}{Black}%
 \GCirc(15,15){4}{0.5}
}}
\makeatletter \@addtoreset{equation}{section} \makeatother

\makeatletter
\renewcommand\section{\@startsection {section}{1}{\z@}%
                                   {-5.5ex \@plus -1ex \@minus -.2ex}% bfr-
                                   {2.3ex \@plus.2ex}%
                                   {\normalfont\large\bfseries}}
\renewcommand\subsection{\@startsection{subsection}{2}{\z@}%
                                     {-3.25ex\@plus -1ex \@minus -.2ex}%
                                     {1.5ex \@plus .2ex}%
                                     {\normalfont\normalsize\bfseries}}
\renewcommand\thesection {\@arabic\c@section}
\renewcommand\thesubsection   {\thesection.\@arabic\c@subsection}
\renewcommand{\@seccntformat}[1]{%
\csname the#1\endcsname.\hspace{1.0em}}
\makeatother
\begin{document}

\flushbottom

%%%%%%%%%%%%%%%%%%%%%%%%%%% TITLE/COVER %%%%%%%%%%%%%%%%%%%%%%%%%%%%%%%%%

\begin{titlepage}

\begin{flushright}
% OUTLINE  \\ 
% DRAFT \\ 
% arXiv:2010.07316\\ 
September 2021
\end{flushright}
\begin{centering}
\vfill

{\Large{\bf
 Mass-suppressed effects in heavy quark diffusion
}} 

\vspace{0.8cm}

A.~Bouttefeux, 
M.~Laine % $^\rmi{a}$
 
\vspace{0.8cm}

% $^\rmi{a}$%
{\em
AEC, 
Institute for Theoretical Physics, 
University of Bern, \\ 
Sidlerstrasse 5, CH-3012 Bern, Switzerland \\}

\vspace*{0.8cm}

\mbox{\bf Abstract}
 
\end{centering}

\vspace*{0.3cm}
 
\noindent
Many lattice studies of heavy quark diffusion
originate from a colour-electric correlator, obtained as a leading term
after an expansion in the inverse of the heavy-quark mass. In view of 
the fact that the charm quark is not particularly heavy, we consider 
subleading terms in the expansion. Working out correlators 
up to $\rmO(1/M^2)$, we argue that the leading corrections 
are suppressed by $\rmO(T/M)$, and one of them 
can be extracted from a colour-magnetic correlator. 
The corresponding transport coefficient is non-perturbative
already at leading order in the weak-coupling expansion, and therefore
requires a non-perturbative determination.  

\vfill

%% %\noindent
%% %PACS numbers: 
%% %11.10.Wx, %        Finite temperature field theory
%% %11.15.Ha, %        Lattice gauge theory  
%% %12.38.Bx, %        Perturbative calculations in QCD
%% %12.38.Mh, %        Quark--gluon plasma
%% %14.40.Nd, %        Bottom mesons
%% %\\
%% %Keywords: Thermal Field Theory, CP violation, Neutrino Physics, Resummation
 
\end{titlepage}

\tableofcontents

%%%%%%%%%%%%%%%%%%%%%%%%%%% SECTION %%%%%%%%%%%%%%%%%%%%%%%%%%%%%%%%%%%%%%
%
\section{Introduction}

The properties of heavy quarks, of mass $M$,  
inserted into a plasma, at a temperature $T$, 
can be characterized by a number of dispersive or mass terms 
($F(\vec{v}) \simeq M^{ }_\rmi{rest} 
 + {M_\rmi{kin} \vec{v}^{2}_{ }} / {2} + ... $)
and absorptive or rate coefficients
(diffusion constant, kinetic and chemical equilibration rates).
In the setting of a heavy ion collision experiment, 
the hierarchy $T \ll M$ is not necessarily drastic, 
particularly for charm quarks. If we determine physical quantities
as a series in $T/M$, it may then be asked how large 
such corrections are, and whether they could help for their part 
to explain the empirical observation that heavy quarks, such as
those identified as $D$~mesons after hadronization, 
appear to interact efficiently with a hot QCD medium~\cite{pheno}. 

In the case of dispersive corrections,
the nature of the series in $T/M$  
is well understood at low orders of  
perturbation theory. Computing thermal effects in 
unresummed perturbation theory leads to a mass correction of
relative magnitude $\rmO(\alphas T^2/M^2)$~\cite{dhr},
however taking into account plasma effects, 
particularly Debye screening, shows that the dominant 
correction is only suppressed by $\rmO(\alphas^{3/2} T/M)$.
For $M^{ }_\rmi{rest}$ this is known as the Salpeter correction 
(cf.,\ e.g.,\ ref.~\cite{lsb}), and a similar effect
also exists for $M_\rmi{kin}$~\cite{chesler}. 
The purpose of the present investigation is to study the
nature of the series for the case of rate coefficients. 

The physics of heavy quark diffusion and kinetic equilibration
is closely related to that of Brownian motion, described by the
Langevin equation. In the non-relativistic limit, this physics
is described by three quantities: the diffusion coefficient, $D$; 
the momentum diffusion coefficient, $\kappa$; and the drag 
coefficient, $\eta$. As already
pointed out by Einstein, these quantities are related to each other
for $T \ll M_\rmi{kin}$, in 
particular $D = 2 T^2/\kappa$ and $\eta = \kappa / (2 M_\rmi{kin}T)$. 
Which of the quantities is viewed as ``primary''  
depends on the context: 
for any mass, $D$ can be expressed
through a Kubo relation which in principle permits for a lattice study; 
in the large-mass limit, $\kappa$ can be expressed
through a Kubo relation which permits for 
a lattice study whose systematic
errors should be better under control than for $D$;
in the large-mass limit,  
$\eta$ can be interpreted 
as a kinetic equilibration rate which leads to a direct physical
interpretation (its inverse can be compared with the medium life time). 

The challenge with a non-perturbative determination of $D$ is that 
the corresponding spectral function shows a very narrow transport
peak in the large-mass and/or weak-coupling limit, 
of width $\eta \sim \alphas^2 T^2/M$~\cite{pt}. 
A controlled reconstruction of 
a spectral function from imaginary-time data is a hard problem, 
and practically impossible in the presence of such sharp features. 
Following a suggestion in ref.~\cite{cst}, the possibility to 
rather extract $\kappa$ was worked out in ref.~\cite{kappaE}. 
In particular, it was argued
that the corresponding spectral function contains 
no sharp transport peak, being instead  
flat at small frequencies. This should allow for 
a somewhat controlled extraction of the transport coefficient, 
and indeed many measurements 
have been carried out in recent years~\cite{lat2,lat25,lat3,lat4,lat5,lat6}, 
supplementing LO~\cite{kappa_lo} and 
NLO~\cite{kappa_nlo} perturbative computations.  

The definition of $\kappa$ in ref.~\cite{kappaE} is related to 
the standard Kubo relation for the diffusion coefficient, $D$. 
Whereas $D$ is obtained as a transport coefficient 
(i.e.\ height of the transport peak) related to 
the 2-point correlator of the vector current
(which is denoted by $\hat{\mathcal{J}}^{ }_i$), 
the idea of ref.~\cite{kappaE} is to instead consider 
the {\em tail} of the transport peak. 
Formally, this is obtained by multiplying
the vector spectral function by $\omega^2$, where $\omega$ is frequency.
Inside a Fourier transform, $\omega$ can be converted 
to a time derivative, so this means that we are really
considering the two-point correlator of an ``acceleration'', 
$
 {\rm d} \hat{\mathcal{J}}^{ }_i / {\rm d}t
$, 
rather than of a ``velocity'', $\hat{\mathcal{J}}^{ }_i$.  
The correct normalization
requires that the whole is multiplied by $M^2_{ }$,\footnote{%
 In ref.~\cite{kappaE} the multiplication was by the thermally 
 corrected $M^{2}_\rmii{kin} = M^2 \,\{ 1 + \rmO(\alphas^{3/2} T/M) \}$, 
 however the $\rmO(\alphas^{3/2} T/M)$ corrections were not treated, 
 so one could have equally multiplied by $M^2$.
 Here we use $M^2$, which does not break Lorentz invariance.
 The overall normalization of $\kappa$ is fixed later while 
 matching onto a low-energy description, 
 via \eqs\nr{einstein} and \nr{Mkin}. 
} 
and divided by the quark number susceptibility, $\chi$, 
which then led ref.~\cite{kappaE} to define 
\be
 \kappa^{\rmii{$(M)$}}_{ }(\omega)
 \; \equiv \; 
 \frac{1}{3\chi} 
 \int_{-\infty}^{\infty}
 \! {\rm d}t \, e^{i\omega t}
 \int_{\vec{x}}
 \Bigl\langle
 \frac{1}{2} 
 \Bigl\{ 
   \hat{\mathcal{F}}^i_{ }(t,\vec{x}\,) \, , \, 
   \hat{\mathcal{F}}^i_{ }(0,\vec{0}\,)
 \Bigr\}  
 \Bigr\rangle
 \;, \la{kappa_M_w}
\ee
where a sum over the spatial indices $i$ is implied, 
and 
$
 \hat{\mathcal{F}}^{ }_i \equiv M {\rm d} 
 \hat{\mathcal{J}}^{ }_i / {\rm d}t
$.
It was shown in ref.~\cite{kappaE} that the ordered limit 
$
 \kappa \equiv \lim_{\omega\to 0}\lim_{M\to \infty}
 \kappa^{\rmii{$(M)$}}_{ }
 (\omega)
$
is an ultraviolet finite observable, 
which can be defined on the non-perturbative level
as a transport coefficient related to the corresponding 
imaginary-time correlator. In the following, 
we take \eq\nr{kappa_M_w} as a starting point, 
and inspect the nature of its $\rmO(T/M)$ corrections. 

This paper is organized as follows. 
We start with a consideration of Lorentz force correlators
within classical electrodynamics (cf.\ \se\ref{se:cl}), 
revealing the key patterns
to be confirmed later on in QCD. 
This is followed by a formal breakdown
of \eq\nr{kappa_M_w} within the large-$M$ expansion
(cf.\ \se\ref{se:derivation}). 
The colour-magnetic correlator emerging from these
considerations is analyzed 
perturbatively (cf.\ \se\ref{se:pert}), 
before we conclude with an outlook (cf.\ \se\ref{se:concl}).

%%%%%%%%%%%%%%%%%%%%%%%%%%%%% SECTION %%%%%%%%%%%%%%%%%%%%%%%%%%%%%%%%%%%%
%
\section{Classical picture}
\la{se:cl}

Consider the Lorentz force acting on 
a probe particle of momentum $\vec{p}$ and charge $q$: 
\be
 \dot{\vec{p}}
 = q \bigl(\vec{E} + \vec{v}\times\vec{B}\, \bigr)(t)
 \; \equiv \; \vec{F}(t)
 \;. \la{lorentz}
\ee
We now imagine a statistical environment, in which the velocities
come from a thermal distribution. 
Given the particle's
large inertia, the time scale of the variation of velocities is 
larger than the time scale of the variation of the electric and
magnetic field strengths. This slow evolution is expected to be 
described by the Langevin equation, 
\be
 \dot{\vec{p}} + \eta \, \vec{p} = \mathbf{f}(t)
 \;, \quad 
 \langle f^{ }_i(t) \rangle = 0 
 \;, \quad 
 \langle f^{ }_i(t') f^{ }_j(t) \rangle = 
 \kappa \, \delta^{ }_{ij} \, \delta(t - t')
 \;. \la{langevin} 
\ee
The dissipative coefficient $\eta$ and the noise self-correlator $\kappa$
are related by the fluctuation-dissipation theorem, following from 
the fact that the solution of the Langevin equation satisfies 
\be
 \lim_{t\to\infty} 
 \langle \vec{p}^{2}(t) \rangle 
 = \frac{3\kappa}{2\eta}
 \;. \la{einstein}
\ee
Furthermore there is a dispersive correction, meaning that the mass
implicit to \eq\nr{langevin}, $M^{ }_\rmi{kin}$, differs from the 
vacuum mass implicit to \eq\nr{lorentz}, $M$. The goal now is 
to extract the ``low-energy parameters'' 
$\kappa$ and $\eta$ from  
properties of the microscopic force in \eq\nr{lorentz}, 
which involves a number of steps, enumerated as follows.  

\paragraph{(i)}
The first step is to compare the right-hand sides of 
\eqs\nr{lorentz} and \nr{langevin}. Given the different time 
scales of evolution, statistical 
averages factorize into averages of velocities 
and averages of fields,\footnote{%
  In order to avoid clutter we denote all statistical averages
  by $\langle ... \rangle$, even if 
  the weight with respect to which the average is taken
  differs from context to context. 
  In point~(ii)
  we return to certain subtleties concerning the circumstances 
  under which the factorization of averages applies. 
   } 
in particular
\be
 \bigl\langle {F}^{ }_i(t') {F}^{ }_j(t) \bigr\rangle
 = q^2\, \biggl\{
   \bigl\langle E^{ }_i(t') E^{ }_j(t) \bigr\rangle 
  + 
  \frac{1}{3} \, 
  \bigl\langle \vec{v}^{2}_{ } \bigr\rangle \,
     \bigl\langle 
        \delta^{ }_{ij} B^{ }_k(t') B^{ }_k(t)
                      - B^{ }_j(t') B^{ }_i(t)
     \bigr\rangle 
  \biggr\} 
  \;.  \la{FF_cl}
\ee 
Matching with \eq\nr{langevin}, $\kappa$ can be extracted as 
a transport coefficient, 
\be
 \kappa = \lim_{\omega\to 0} \int_{-\infty}^{\infty} \! {\rm d}t' 
 \, e^{i\omega (t' - t)}
 \frac{1}{3} \sum_i \langle F^{ }_i(t') F^{ }_i(t) \rangle
 \;. \la{kubo}
\ee
According to \eq\nr{FF_cl}, 
the leading term originates from the colour-electric correlator~\cite{cst}, 
and the first correction from a colour-magnetic one, 
whose contribution is suppressed
by $ \langle \vec{v}^{2}_{ }\rangle \sim \rmO(T/M^{ }_{ })$ 
according to \eq\nr{Mkin}. 

\paragraph{(ii)}
Actually, the argument above is a bit sloppy, as the spatial positions
of the fields are suppressed. In reality, if a heavy particle starts 
from $\vec{x} =\vec{0}$ at time $t=0$ and has the velocity 
$\vec{v}$, at a later time $t$ it is at 
$\vec{x}(t) = \vec{v} t + \rmO(\dot{\vec{v}} t^2)$.   
Then we should really insert fields as 
$
 E^{ }_i (t,\vec{x}(t)) = 
 E^{ }_i(t,\vec{0}) + t\, v^{ }_j \partial^{ }_j E^{ }_i(t,\vec{0}) 
 + \ldots
$, 
and similarly for 
$
 B^{ }_i (t,\vec{x}(t)) 
$.
For instance, 
when correlated with $\vec{v}\times\vec{B}$, 
the next-to-leading terms in the expansion of $E^{ }_i$ 
lead to further effects of $\langle \vec{v}^{2}_{ }\rangle$, 
however multiplied by powers of $t$. 
Such corrections may be called {\em secular terms}. Any practical
study should be formulated such that secular terms are avoided,\footnote{%
  In a Langevin simulation
  (cf.,\ e.g.,\ ref.~\cite{kappa_lo}), 
  this is done by tracking separately
  the positions and momenta of the heavy quarks: 
  ${\rm d} \vec{x} = \vec{v}{\rm d} t$, 
  ${\rm d} \vec{p} = (-\eta\, \vec{p} + \vec{f}){\rm d} t$, 
  with $\vec{v} = \vec{p} / \sqrt{\vec{p}^2 + M_\rmii{kin}^2}$.
  In this case, $\vec{f}$ can be 
  generated from a random 
  ensemble applicable to 
  the ``old'' position $\vec{x}$. \la{secular}
 } 
and we have to watch out for them in \se\ref{se:derivation} as well. 

\paragraph{(iii)}
There is a further subtlety, related to the difference between
non-relativistic and relativistic momenta. 
Suppose that we consider a non-relativistic momentum 
$M^{ }_{ } \vec{v}$ rather than a relativistic one, 
$\vec{p} \equiv M^{ }_{ } \vec{u}$, with $\vec{u} \equiv \gamma \vec{v}$
(the reason for this should become
apparent around \eq\nr{Ji_cl}).
Writing $\vec{v}  =\gamma^{-1} \vec{u}$ we find
\be
 \dot{v}^{ }_i = \gamma^{-1} \dot{u}^{ }_i - u^{ }_i u^{ }_j \dot{v}^{ }_j
 \;. 
\ee
Moving the last term to the left-hand side, 
and solving the matrix equation for $\dot{v}^{ }_i$, 
leads to 
\be
 \dot{v}^{ }_i = \bigl(\delta^{ }_{ij} - v^{ }_i v^{ }_j \bigr)
 \gamma^{-1} \dot{u}^{ }_j
 \;. \la{acc1}
\ee
Taking the 2-point correlator of these accelerations, and 
expanding to second order in small velocities, gives
\be
 M^2_{ } \bigl\langle \dot{v}^{ }_i(t') \dot{v}^{ }_j(t) \bigr\rangle
 = 
 \bigl\langle {F}^{ }_i(t') {F}^{ }_j(t) \bigr\rangle
 - 
 \bigl\langle \vec{v}^{2}_{ } \delta^{ }_{ik} \delta^{ }_{jl} 
  + v^{ }_i v^{ }_k \delta^{ }_{jl} 
  + \delta^{ }_{i k} v^{ }_j v^{ }_l \bigr\rangle
 \bigl\langle {F}^{ }_k(t') {F}^{ }_l(t) \bigr\rangle
 + 
 \rmO(v^4)
 \;. \la{acc2}
\ee
Here the force-force correlator 
can be inserted from \eq\nr{FF_cl}.
In order to avoid the second term in \eq\nr{acc2}, 
we should use relativistic momenta. 

\paragraph{(iv)}
Finally, we have to consider the relation of $\kappa$ and $\eta$, 
originating from \eq\nr{einstein}. 
Equipartition in classical statistical physics implies
$
 \langle p^{ }_i \partial H / \partial p^{ }_i \rangle = T
$, 
where $H$ is the Hamiltonian and no sum over $i$ is taken. 
For a 1-particle Hamiltonian this corresponds to 
$
 \langle \vec{p}\cdot\vec{v} \rangle = 3 T 
$.
Inserting a covariant momentum, with the mass 
including a dispersive correction as we are now within the 
low-energy effective description, leads to 
\be
 \langle \gamma \vec{v}^2 \rangle = \frac{3 T}{M^{ }_\rmii{kin}} 
 \;. \la{Mkin}
\ee
Up to next-to-leading order in velocities, we may expand 
$ 
 \langle \gamma \vec{v}^2 \rangle
 \approx 
 \langle
   \vec{v}^2 + (\vec{v}^2)^2/2 
 \rangle
$.
For the quartic part, we can take a Gaussian average, 
$
 \langle (\vec{v}^2)^2 \rangle \approx \frac{5}{3} 
 \langle \vec{v}^2 \rangle^2
$.
Inserting into \eq\nr{Mkin}, leads to 
$
 \langle \vec{v}^2 \rangle 
 \approx ({3T}/ {M_\rmi{kin}})
 \{  
  1 - 5 T / (2 M_\rmi{kin})
 \}
$,\footnote{%
  At leading order in $\alphas$, $\langle\vec{v}^2\rangle$ 
  can also be defined as the area under the transport peak 
  in the correlator 
  $\langle \hat{\mathcal{J}}^{ }_i \hat{\mathcal{J}}^{ }_i \rangle$,
  normalized to the susceptibility $\chi$~\cite{kappaE}. 
  From \eqs(3.4), (3.5) of ref.~\cite{GVtau}, this leads to 
  $\langle\vec{v}^2\rangle = \int_{\vec{p}} (p/E^{ }_p)^2 \nF'(E^{ }_p)
                           / \int_{\vec{p}} \nF'(E^{ }_p) $,
  where $E^{ }_p = \sqrt{p^2 + M^2}$
  and $\nF^{ }$ is the Fermi distribution. After an expansion in $T/M$
  up to next-to-leading order, 
  this agrees with the result we cite here.
 } 
and this then gives 
$
 \langle \vec{p}^2 \rangle 
 \approx 
 3 M_\rmi{kin} T 
 \{  
  1 + 5 T / (2 M_\rmi{kin})
 \}
$.
In combination with \eq\nr{einstein}, we finally find 
\be
 \eta \approx \frac{\kappa}{2 M_\rmii{kin} T}
 \,\biggl(
  1 - \frac{5 T}{2 M_\rmii{kin}} 
 \biggr)
 \;. \la{einstein2}
\ee

\paragraph{}
To summarize, the force felt by 
a non-relativistic probe particle  
experiences four types of corrections\footnote{%  
  It is unclear to us whether the corrections discussed
  could be related to  
  a generalization of the Langevin equation used for describing heavy
  quarks having relativistic (i.e.\ non-equilibrated) momenta with respect 
  to the medium. It has been suggested that in this case 
  the ``comoving'' forces could be 
  correlated as~\cite{kappa_lo} 
  \be
   \langle f^{ }_i(t') f^{ }_j(t) \rangle 
   \; \simeq \;
   \bigl\{ \, 
     \bigl( \delta^{ }_{ij} - \hat{p}^{ }_i\,\hat{p}^{ }_j \bigr) 
     \, \kappa^{ }_\rmiii{T}(p) 
    + 
     \hat{p}^{ }_i\,\hat{p}^{ }_j
     \, \kappa^{ }_\rmiii{L}(p) 
   \bigr\} \,  
   \delta(t-t')
   \;, 
  \ee
  where $\hat{p}^{ }_i \equiv p^{ }_i / {p}$.
  In our equilibrated case, considering local forces 
  as described in footnote~\ref{secular}, 
  the correlator
  is proportional to $\delta^{ }_{ij}$, implying effectively that 
  $\kappa^{ }_\rmiii{L}(p) = \kappa^{ }_\rmiii{T}(p) = \kappa$,
  where $\kappa$ includes corrections of $\rmO(T/M)$.
  }  
at first order in $\langle \vec{v}^{2}_{ }\rangle \ll 1$.
Summing over the indices, 
the first correction is from 
$
      \bigl\langle B^{ }_i(t') B^{ }_i(t) \bigr\rangle 
$, 
weighted by
$
 \frac{2}{3} \langle \vec{v}^{2}_{ }\rangle
$
according to \eq\nr{FF_cl}. 
The second correction are secular terms, of the type discussed 
under point (ii).
The third correction is from 
$
      \bigl\langle E^{ }_k(t') E^{ }_l(t) \bigr\rangle 
$, 
weighted by 
$
 - \bigl\langle \vec{v}^{2}_{ } \delta^{ }_{ik} \delta^{ }_{jl} 
  + v^{ }_i v^{ }_k \delta^{ }_{jl} 
  + \delta^{ }_{i k} v^{ }_j v^{ }_l \bigr\rangle
 = - \frac{5}{3} \langle \vec{v}^{2}_{ }\rangle
     \delta^{ }_{ik} \delta^{ }_{jl} 
$
according to \eq\nr{acc2}. This a ``trivial'' effect, 
inhibiting acceleration towards the speed of light, 
and eliminated simply by going from 
$M^{ }_{ }\vec{v}$ back into 
the covariant momentum $\vec{p}$.
The fourth correction is to the fluctuation-dissipation 
relation, according to \eq\nr{einstein2}. 
Finally, there is a dispersive effect, substituting
the vacuum mass $M$ through a thermally corrected $M_\rmi{kin}$.

%%%%%%%%%%%%%%%%%%%%%%%%%%% SECTION %%%%%%%%%%%%%%%%%%%%%%%%%%%%%%%%%%%%%%
%
\section{Formal derivation of the force-force correlator}
\la{se:derivation}

Motivated by the discussion of the previous section and in
particular \eq\nr{FF_cl}, the goal now
is to derive an expression for the ``microscopic'' 
force-force correlator in QCD. 

%%%%%%%%%%%%%%%%%%%%%%%%%%% SUBSECTION %%%%%%%%%%%%%%%%%%%%%%%%%%%%%%%%%%%
%
\subsection{Action up to $\rmO(1/M^3)$}
\la{ss:SM}

Let $\theta$ be a $2\Nc^{ }$-component non-relativistic spinor; 
$g$ a gauge coupling; 
$D^{ }_\mu \equiv \partial^{ }_\mu - i g A^{ }_\mu$ a covariant derivative; 
$g E^{ }_i \equiv i [D^{ }_0, D^{ }_i]$ a colour-electric field; 
$g B^{ }_i \equiv \frac{i}{2} \epsilon^{ }_{ijk}[D^{ }_j,D^{ }_k]$
a colour-magnetic field; 
and $\sigma^{ }_i$ the Pauli matrices. 
Starting from the Minkowskian QCD action for one heavy quark flavour, 
$ S^{ }_\M = \int_\mathcal{X} 
  \bar\psi (i \gamma^{\mu}_{ }D^{ }_\mu - M ) \psi
$, 
where $\int_\mathcal{X} \equiv \int \!{\rm d}t \int\!{\rm d}^3\vec{x}$,  
and carrying out a standard tree-level computation, yields 
the expansion
\ba
 S^{ }_\M & \supset & \int_\mathcal{X}
 \, \theta^\dagger \, \biggl\{ \, 
 i D^{ }_0 - M + \frac{\vec{D}^2 + g \vec{\sigma}\cdot\vec{B}}{2M}
 - \frac{g \bigl[ \vec{D}\cdot\vec{E} \bigr] 
   + i g \vec{\sigma}\cdot 
    \bigl( \vec{D}\times\vec{E} - \vec{E}\times\vec{D} \bigr)}{8 M^2}
 \nn 
 & + & 
 \frac{ 
  \vec{D}^4_{ }
  + g \bigl\{ \vec{D}^2,\vec{\sigma}\cdot\vec{B} \bigr\} 
  + g^2 \bigl(\vec{B}^2 - \vec{E}^2\bigr) 
  + i g^2 \vec{\sigma}\cdot
    \bigl( \vec{B}\times\vec{B} - \vec{E}\times\vec{E} \bigr)
 }{8M^3} 
 + \rmO\biggl( \frac{1}{M^4} \biggr)
 \biggr\}\, \theta
 \;, \hspace*{5mm} \nn \la{SM}
\ea
where the antiparticle part has been omitted. 
At the quantum level,  
further operators get generated
and the coefficients of the operators get corrected~\cite{manohar}.
Such results are of no direct use to us, however, given that 
the time derivative in \se\ref{ss:F} is a short-distance 
operation and thus changes quantum corrections. We note, furthermore,
that the rest mass is often omitted from \eq\nr{SM}, but we keep it visible 
as a residual mass, as the corresponding Boltzmann factor plays
an important role at finite temperature. With these specifications,
\eq\nr{SM} serves as the starting point of our investigation. 

%%%%%%%%%%%%%%%%%%%%%%%%%%% SUBSECTION %%%%%%%%%%%%%%%%%%%%%%%%%%%%%%%%%%%
%
\subsection{Lorentz force up to $\rmO(1/M^2)$}
\la{ss:F}

The goal now is to derive the Lorentz force originating from \eq\nr{SM}. 
Viewing $\theta$ and $\theta^*$ as independent fields, the Noether current
can be defined as 
\be
 \mathcal{J}^{\mu}_{ } \; \equiv \; 
 i \biggl[
   \theta^{*}_{\alpha r} 
     \frac{\delta}{\delta(\partial_\mu\theta^{*}_{\alpha r})}
 - \theta^{ }_{\alpha r} 
     \frac{\delta}{\delta(\partial_\mu\theta^{ }_{\alpha r})}
 \biggr] \, S^{ }_\M
 \;, \la{Jmu}
\ee
where $\alpha \in \{1,...,\Nc\}$ labels 
colour and $r \in \{1,2\}$ spin. 
This yields the components 
\ba
 \mathcal{J}^0_{ } & = & \theta^\dagger \theta
 \;, \la{J0}
 \\ 
%%%%%%
 \mathcal{J}^i_{ } & = &  
 \frac{i \theta^\dagger\bigl( \overleftarrow{D}^{ }_{\!i} - 
  \overrightarrow{D\,}^{ }_{\!i} \bigr) \theta}{2M}
 + \frac{g \epsilon^{ }_{ijk}
     \theta^\dagger \sigma^{ }_j E^{ }_k \theta}{4M^2}
% \nn  & - &
   -  \frac{i \theta^\dagger 
            \bigl\{ D^{ }_i,\vec{D}^2 + g \vec{\sigma}\cdot \vec{B}
            \bigr\} \theta }
            {4 M^3}
   + \rmO\Bigl(\frac{1}{M^4} \Bigr)
  \;. \la{Ji}
\ea

Before proceeding, let us note 
that if we consider a spinless 1-particle plane wave, 
$\theta \simeq Z \, e^{-iEt + i\vec{p}\cdot\vec{x}}$, 
$\theta^\dagger \simeq Z^\dagger \, e^{iEt-i\vec{p}\cdot\vec{x}}$, 
then the on-shell point of \eq\nr{SM} corresponds to the relation 
$E = M + \frac{p^2}{2M} - \frac{p^4}{8M^3} + ... = \sqrt{p^2+M^2}$.
The Noether charge density in \eq\nr{J0} evaluates to 
$\mathcal{J}^0_{ } \simeq |Z|^2$, and the current in \eq\nr{Ji} to 
\be
 \mathcal{J}^i_{ } \simeq |Z|^2 \, 
 \biggl( \frac{p_i}{M} - \frac{p_i p^2}{2 M^3} + ... \biggr) 
 = \frac{|Z|^2 p_i}{\sqrt{p^2 + M^2}}
 = |Z|^2 v^{ }_i 
 \;. \la{Ji_cl}
\ee
Therefore, $\mathcal{J}^i_{ }/\mathcal{J}^0_{ }$ 
represents the velocity rather than 
the covariant velocity, and we expect 
the argument around \eqs\nr{acc1} and \nr{acc2} to apply to 
considerations following from \eq\nr{kappa_M_w}. 

To obtain the Lorentz force, we take a time derivative
of the current, and use equations of motion. As the observable
in \eq\nr{kappa_M_w} involves a spatial average, 
partial integrations are permitted in spatial directions. 
Obviously the Noether charge 
$\int_{\vec{x}} \mathcal{J}^0$ is conserved, 
but the spatial currents are not, obeying  
\ba
 \int_{\vec{x}} M \partial^{ }_0 \mathcal{J}^{i}_{ } & = & 
 \int_{\vec{x}} \theta^\dagger \, \biggl\{ 
 - g E^{ }_i 
 + \frac{\bigl[ D^{ }_i,\vec{D}^2 + g \vec{\sigma}\cdot\vec{B} \,\bigr]}{2M}
 + \frac{g\bigl[ D^{ }_0,\vec{\sigma}\times\vec{E} \,\bigr]^{ }_i }{4M}
%%%%%%%%%%
 \nn 
 & - & 
 \frac{g \bigl[ D^{ }_i, [\vec{D}\cdot\vec{E}] + 
  i \vec{\sigma}\cdot(\vec{D}\times\vec{E}
     - \vec{E}\times\vec{D}) \,\bigr]
 + ig \bigl[
     \vec{D}^2 + g\vec{\sigma}\cdot\vec{B} ,  
     \vec{\sigma}\times\vec{E}
    \,\bigr]^{ }_i }{8M^2}
%%%%%%%%%%
 \nn 
 & - &  
 \frac{
  i \bigl[ D^{ }_0 , 
  \{ D^{ }_i, \vec{D}^2 + g\vec{\sigma}\cdot\vec{B}  \}\,\bigr]
      }{4M^2}
 + \rmO\Bigl(\frac{1}{M^3} \Bigr)
 \biggr\}\,\theta
 \;. \la{d0Ji}
\ea
Let us stress again that at the quantum level, 
further operators and non-trivial Wilson coefficients 
are generated, but here we remain at the tree level.  

In view of a lattice study or thermal field theory computation, 
the final step is to Wick rotate the result to Euclidean signature:
$D^{ }_0 \to i D^{ }_0$, 
$E^{ }_i \to i E^{ }_i$.
Furthermore we reduce the number of explicit derivatives
appearing in \eq\nr{d0Ji}, by making use of 
\ba
 \bigl[ D^{ }_i,\vec{D}^2 \, \bigr]
 \!\! & = & \!\!\! 
 -i g \bigl( \vec{D}\times\vec{B} - \vec{B}\times\vec{D} \bigr)^{ }_i
 \;, \\ 
 \bigl[ D^{ }_0, \bigl\{ D^{ }_i,\vec{D}^2 \, \bigr\} \bigr]
 \!\! & = & \!\!\!
  -i g \bigl(
          E^{ }_i \overrightarrow{D\,}^{\!2}
        + \overleftarrow{D\,}^{\!2} E^{ }_i 
        + 
              E^{ }_j
              \overrightarrow{D\,}^{ }_{\!j} 
              \overrightarrow{D\,}^{ }_{\!i}
        + 
              \overleftarrow{D\,}^{ }_{\!i}
              \overleftarrow{D\,}^{ }_{\!j} 
              E^{ }_j 
        - 
              \overleftarrow{D\,}^{ }_{\!j} 
              E^{ }_j 
              \overrightarrow{D\,}^{ }_{\!i}
        - 
              \overleftarrow{D\,}^{ }_{\!i} 
              E^{ }_j 
              \overrightarrow{D\,}^{ }_{\!j}
        \bigr)
  \;. \hspace*{6mm}
\ea
Lowering the index $i$ on the left-hand side of \eq\nr{d0Ji}, 
in order to insert an overall minus sign, 
but suppressing it from the notation, 
the Lorentz force operators 
originating from \eq\nr{d0Ji} can then be enumerated as  
\ba
 O^{ }_0 & \equiv & \theta^\dagger i g E^{ }_i \theta 
 \;, \la{O0} \\[2mm] 
 O^{ }_{1a} & \equiv & \frac{\theta^\dagger 
   i g (\vec{D}\times\vec{B} - \vec{B}\times\vec{D})^{ }_i \theta}{2M}
 \;, \la{O1a} \\ 
 O^{ }_{1b} & \equiv & \frac{\theta^\dagger
   (-g)[D^{ }_i,\vec{\sigma}\cdot\vec{B}\,]\,\theta}{2M}
 \;, \la{O1b} \\ 
 O^{ }_{1c} & \equiv & \frac{\theta^\dagger 
   g[D^{ }_0,\vec{\sigma}\times\vec{E}\,]^{ }_i \,\theta}{4M}
 \;, \la{O1c} \\ 
 O^{ }_{2a} & \equiv & \frac{\theta^\dagger
   ig [D^{ }_i,[\vec{D}\cdot\vec{E}\,]]
   \,\theta}{8M^2}
 \;, \la{O2a} \\ 
 O^{ }_{2b} & \equiv & \frac{\theta^\dagger
   (-g) [D^{ }_i, \vec{\sigma}\cdot( \vec{D}\times\vec{E}
                                    -\vec{E}\times\vec{D} )\,]
   \,\theta}{8M^2}
 \;, \la{O2b} \\ 
 O^{ }_{2c} & \equiv & \frac{\theta^\dagger
   g [ \vec{\sigma} \times\vec{E}, 
       \vec{D}^2 \,]^{ }_i 
   \,\theta}{8M^2}
 \;, \la{O2c} \\ 
 O^{ }_{2d} & \equiv & \frac{\theta^\dagger
   g^2 [ \vec{\sigma} \times\vec{E}, 
       \sigma\cdot\vec{B} \,]^{ }_i 
   \,\theta}{8M^2}
 \;, \la{O2d} \\ 
 O^{ }_{2e} & \equiv & \frac{\theta^\dagger
    ig (
          E^{ }_i \overrightarrow{D\,}^{\!2}
        + \overleftarrow{D\,}^{\!2} E^{ }_i 
        + 
              E^{ }_j
              \overrightarrow{D\,}^{ }_{\!j} 
              \overrightarrow{D\,}^{ }_{\!i}
        + 
              \overleftarrow{D\,}^{ }_{\!i}
              \overleftarrow{D\,}^{ }_{\!j} 
              E^{ }_j 
        - 
              \overleftarrow{D\,}^{ }_{\!j} 
              E^{ }_j 
              \overrightarrow{D\,}^{ }_{\!i}
        - 
              \overleftarrow{D\,}^{ }_{\!i} 
              E^{ }_j 
              \overrightarrow{D\,}^{ }_{\!j}
        )
   \,\theta}{4M^2}
 \;, \la{O2e} \hspace*{5mm} \\
 O^{ }_{2f} & \equiv & \frac{\theta^\dagger
   (-g) [ D^{ }_0,\{D^{ }_i,\vec{\sigma}\cdot\vec{B} \} \,]
   \,\theta}{4M^2}
 \;. \la{O2f} 
\ea
The Euclidean action with respect to which thermal averages 
are taken ($e^{i S^{ }_\M} \to e^{ -S^{ }_\E}$) becomes
\ba
 S^{ }_\E & = & S^{ }_0 + S^{ }_1 + S^{ }_2 + \rmO\Bigl(\frac{1}{M^3}\Bigr)
 \;, \\ 
 S^{ }_0 & = & \int_X 
 \theta^\dagger \,\bigl( D^{ }_0 + M \,\bigr) \, \theta
 \;, \la{S0} \\ 
 S^{ }_1 & = & - \int_X
 \frac{\theta^\dagger \,\bigl( \vec{D}^2
               + g \vec{\sigma}\cdot\vec{B} \,\bigr) \, \theta}{2M}
 \;, \la{S1} \\ 
 S^{ }_2 & = &    \int_X
 \frac{\theta^\dagger (ig) \bigl[\vec{D}\cdot\vec{E}\,\bigr] \theta 
      - \theta^\dagger \, g \vec{\sigma}\cdot
              \bigl( \vec{D}\times\vec{E} 
                   - \vec{E}\times\vec{D}  \,\bigr) \, \theta}{8M^2}
 \;, \la{S2}
\ea
where 
$
 \int_X \equiv \int_0^\beta \! {\rm d}\tau \int_{\vec{x}} 
$.
We note that in normal HQET, 
only $S^{ }_0$ appears as a weight in the thermal average, 
whereas $S^{ }_1$ and $S^{ }_2$ are expanded into 
correlation functions, however in the thermal context 
taking this limit 
requires care, as will be discussed in \se\ref{ss:BB}.

%%%%%%%%%%%%%%%%%%%%%%%%%%% SUBSECTION %%%%%%%%%%%%%%%%%%%%%%%%%%%%%%%%%%%
%
\subsection{Force-force correlator up to $\rmO(1/M^2)$}
\la{ss:FF}

Let us consider the 2-point correlation function of the operator
defined as a sum of \eqs\nr{O0}--\nr{O2f}. 
As alluded to at the end of \se\ref{ss:F}, 
we do not expand the action in powers of $1/M$ yet, 
i.e.\ we are working within the NRQCD rather than HQET action for a moment. 

After inserting the operators, we carry out Wick contractions
for the fermion fields. 
A fermion propagator (which is a $2\Nc^{ } \times 2\Nc^{ }$-matrix) 
is defined as 
\be
 \Delta^{ }_{ }(\tau^{ }_2,\vec{y};\tau^{ }_1,\vec{x}\,)
 \; \equiv \; 
 \bigl\langle \,
  \theta^{ }_{ }(\tau^{ }_2,\vec{y}\,)\, 
  \theta^{\dagger}_{ }(\tau^{ }_1,\vec{x}\,) 
 \, \bigr\rangle
 \;, 
\ee
and covariant derivatives acting on it as
\be
% \{ 
 \overrightarrow{D\,}^{ }_{\!i}
 \overrightarrow{D\,}^{ }_{\!j} ... 
 \Delta (\tau^{ }_2,\vec{y};\tau^{ }_1,\vec{x}\,) 
  ...
 \overleftarrow{D}^{ }_{\!k} 
 \overleftarrow{D}^{ }_{\!l}
% \}^{ }_{\alpha\beta;rs} 
 \; \equiv \; 
 \Delta^{ }_{ij...;...kl} % _{\alpha\beta;rs}
 (\tau^{ }_2,\vec{y};\tau^{ }_1,\vec{x}\,)
 \;. 
\ee
Correlators are denoted by 
\be
 G^{ }_{mn}(\tau) \; \equiv \; 
 \frac{
 \sum_{i=1}^{3}
 \int_{\vec{x}} 
 \bigl\langle
   O^{ }_{m}(\tau,\vec{x}\,) O^{ }_{n}(0,\vec{0}\,)
 \bigr\rangle}{3\chi} 
 \;, \quad
 G^{ }_{\{mn\}}(\tau) \; \equiv \; 
% \frac{ 
  G^{ }_{mn}(\tau) + G^{ }_{nm}(\tau)
% }{2}
 \;, 
\ee
where the quark number susceptibility reads 
\be
 \chi \; \equiv \; \int_{\vec{x}}
  \bigl\langle
   \theta^\dagger\theta(\tau,\vec{x}\,) \,
   \theta^\dagger\theta(0,\vec{0}\,)
 \bigr\rangle
 \;, \quad \tau > 0 
 \;. 
 \la{chi_def}
\ee
The value of $\chi$ is independent
of the choice of $\tau$, because $\int_{\vec{x}} \theta^\dagger\theta$ 
is a conserved charge within the NRQCD/HQET action. 

At leading order in the $1/M$-expansion, 
i.e.\ with the operator from \eq\nr{O0}, 
the definitions above lead to 
\be
 G^{ }_{00}(\tau) = 
 - \frac{g^2}{3\chi} 
%  \sum_{i=1}^{3}
 \int_{\vec{x}} 
 \tr 
 \bigl\langle
   \Delta(\beta,\vec{0};\tau,\vec{x}\,) E^{ }_i(\tau,\vec{x}\,) 
   \Delta(\tau,\vec{x};0,\vec{0}\,) E^{ }_i(0,\vec{0}\,)
 \bigr\rangle
 \;, \la{G_00}
\ee
where a sum over the index $i$ is implied. 

Proceeding to $\rmO(1/M)$, we are faced with the correlators 
$G^{ }_{\{01a\}}$,  
$G^{ }_{\{01b\}}$, and   
$G^{ }_{\{01c\}}$.  
Given that the operators $O^{ }_{1b}$ and $O^{ }_{1c}$ contain 
a Pauli matrix and that spin effects only appear at $\rmO(1/M)$
in \eq\nr{S1}, the latter two vanish up to $\rmO(1/M^2)$. 
The remaining one gives
\ba
 G^{ }_{\{01a\}}(\tau) & = &  
 \frac{g^2  \epsilon^{ }_{ijk}
 }{6 M \chi} 
% \sum_{i,j,k}
 \int_{\vec{x}} 
 \tr 
 \bigl\langle
   \Delta^{ }_{ }(\beta,\vec{0};\tau,\vec{x}\,) E^{ }_i(\tau,\vec{x}\,) 
   \Delta^{ }_{;j}(\tau,\vec{x};0,\vec{0}\,) B^{ }_k(0,\vec{0}\,)
%%%%%%
 \nn[0mm] 
 & & \hspace*{1.8cm} 
   - \, 
   \Delta^{ }_{j;}(\beta,\vec{0};\tau,\vec{x}\,) E^{ }_i(\tau,\vec{x}\,) 
   \Delta^{ }_{ }(\tau,\vec{x};0,\vec{0}\,) B^{ }_k(0,\vec{0}\,)
%%%%%%
 \nn[2mm] 
 & & \hspace*{1.8cm} 
   + \, 
   \Delta^{ }_{;j}(\beta,\vec{0};\tau,\vec{x}\,) B^{ }_k(\tau,\vec{x}\,) 
   \Delta^{ }_{ }(\tau,\vec{x};0,\vec{0}\,) E^{ }_i(0,\vec{0}\,)
%%%%%%
 \nn[2mm] 
 & & \hspace*{1.8cm} 
   - \, 
   \Delta^{ }_{ }(\beta,\vec{0};\tau,\vec{x}\,) B^{ }_k(\tau,\vec{x}\,) 
   \Delta^{ }_{j;}(\tau,\vec{x};0,\vec{0}\,) E^{ }_i(0,\vec{0}\,)
 \; 
 \bigr\rangle
 \;. \hspace*{5mm} \la{G01a}
\ea
We return to a discussion of this correlator in \se\ref{ss:BB}.

Finally, at order $1/M^2$, the correlators are 
$G^{ }_{1a1a}$, 
$G^{ }_{1b1b}$, 
$G^{ }_{1c1c}$, 
$G^{ }_{\{1a1b\}}$, 
$G^{ }_{\{1a1c\}}$, 
$G^{ }_{\{1b1c\}}$, 
$G^{ }_{\{02a\}}$, 
$G^{ }_{\{02b\}}$, 
$G^{ }_{\{02c\}}$, 
$G^{ }_{\{02d\}}$, 
$G^{ }_{\{02e\}}$, and  
$G^{ }_{\{02f\}}$.
Among these,  
$G^{ }_{\{1a1b\}}$, 
$G^{ }_{\{1a1c\}}$, 
$G^{ }_{\{02b\}}$, 
$G^{ }_{\{02c\}}$, and  
$G^{ }_{\{02f\}}$
are really of $\rmO(1/M^3)$, because only one of 
the operators contains a Pauli matrix, and spin-dependent effects 
are suppressed by $1/M$ in the action, cf.\ \eq\nr{S1}. 
Among the rest, 
$G^{ }_{1b1b}$, 
$G^{ }_{1c1c}$, 
$G^{ }_{\{1b1c\}}$, 
$G^{ }_{\{02a\}}$, and 
$G^{ }_{\{02d\}}$ 
are of $\rmO(1/M^2)$.
The two remaining ones, 
$G^{ }_{1a1a}$ and 
$G^{ }_{\{02e\}}$, 
are the most important ones, as they contain
derivatives acting on the heavy quark propagators that cannot be 
eliminated by partial integrations. As discussed in \se\ref{ss:BB},  
these two correlators are really of $\rmO(T/M)$, and deserve to 
be given explicitly: 
\ba
 G^{ }_{1a1a}(\tau) & = &  
 -\frac{g^2 
 \epsilon^{ }_{ijk}
 \epsilon^{ }_{imn}
  }{12 M^2 \chi} 
% \sum_{i,j,k,m,n}
 \int_{\vec{x}} 
 \tr 
 \bigl\langle
   \Delta^{ }_{;j}(\beta,\vec{0};\tau,\vec{x}\,) B^{ }_k(\tau,\vec{x}\,) 
   \Delta^{ }_{;m}(\tau,\vec{x};0,\vec{0}\,) B^{ }_n(0,\vec{0}\,)
%%%%%%
 \nn[0mm] 
 & & \hspace*{2.8cm} 
   + \, 
   \Delta^{ }_{n;j}(\beta,\vec{0};\tau,\vec{x}\,) B^{ }_k(\tau,\vec{x}\,) 
   \Delta^{ }_{ }(\tau,\vec{x};0,\vec{0}\,) B^{ }_m(0,\vec{0}\,)
%%%%%%
 \nn[2mm] 
 & & \hspace*{2.8cm} 
   + \, 
   \Delta^{ }_{ }(\beta,\vec{0};\tau,\vec{x}\,) B^{ }_j(\tau,\vec{x}\,) 
   \Delta^{ }_{k;m}(\tau,\vec{x};0,\vec{0}\,) B^{ }_n(0,\vec{0}\,)
%%%%%%
 \nn[2mm] 
 & & \hspace*{2.8cm} 
   + \, 
   \Delta^{ }_{n;}(\beta,\vec{0};\tau,\vec{x}\,) B^{ }_j(\tau,\vec{x}\,) 
   \Delta^{ }_{k;}(\tau,\vec{x};0,\vec{0}\,) B^{ }_m(0,\vec{0}\,)
 \; 
 \bigr\rangle
 \;, \hspace*{7mm} \la{G1a1a} \\[3mm]
%%%%%%%%%%%%%%%%%%%%%%%%%%%%%%%%%%%%%%
 G^{ }_{02a}(\tau) & = &  
 - \frac{g^2 
 }{12 M^2 \chi} 
% \sum_{i,j,k}
 \int_{\vec{x}} 
 \tr 
 \bigl\langle
   \Delta^{ }_{jj;}(\beta,\vec{0};\tau,\vec{x}\,) E^{ }_i(\tau,\vec{x}\,) 
   \Delta^{ }_{ }(\tau,\vec{x};0,\vec{0}\,) E^{ }_i(0,\vec{0}\,)
%%%%%%
 \nn[0mm] 
 & & \hspace*{2.4cm} 
   + \, 
   \Delta^{ }_{ }(\beta,\vec{0};\tau,\vec{x}\,) E^{ }_i(\tau,\vec{x}\,) 
   \Delta^{ }_{;jj}(\tau,\vec{x};0,\vec{0}\,) E^{ }_i(0,\vec{0}\,)
%%%%%%
 \nn[2mm] 
 & & \hspace*{2.4cm} 
   + \, 
   \Delta^{ }_{ji;}(\beta,\vec{0};\tau,\vec{x}\,) E^{ }_i(\tau,\vec{x}\,) 
   \Delta^{ }_{ }(\tau,\vec{x};0,\vec{0}\,) E^{ }_j(0,\vec{0}\,)
%%%%%%
 \nn[2mm] 
 & & \hspace*{2.4cm} 
   + \, 
   \Delta^{ }_{ }(\beta,\vec{0};\tau,\vec{x}\,) E^{ }_i(\tau,\vec{x}\,) 
   \Delta^{ }_{;ij}(\tau,\vec{x};0,\vec{0}\,) E^{ }_j(0,\vec{0}\,)
%%%%%%
 \nn[2mm] 
 & & \hspace*{2.4cm} 
   - \, 
   \Delta^{ }_{i;}(\beta,\vec{0};\tau,\vec{x}\,) E^{ }_i(\tau,\vec{x}\,) 
   \Delta^{ }_{;j}(\tau,\vec{x};0,\vec{0}\,) E^{ }_j(0,\vec{0}\,)
%%%%%%
 \nn[2mm] 
 & & \hspace*{2.4cm} 
   - \, 
   \Delta^{ }_{j;}(\beta,\vec{0};\tau,\vec{x}\,) E^{ }_i(\tau,\vec{x}\,) 
   \Delta^{ }_{;i}(\tau,\vec{x};0,\vec{0}\,) E^{ }_j(0,\vec{0}\,)
 \; 
 \bigr\rangle
 \;. \hspace*{5mm} \la{G02a}
\ea
In \eq\nr{G02a} we have displayed only one ordering, 
with the other one in 
$G^{ }_{\{02a\}} = G^{ }_{02a} + G^{ }_{2a0}$
giving a similar result. 

%%%%%%%%%%%%%%%%%%%%%%%%%%% SUBSECTION %%%%%%%%%%%%%%%%%%%%%%%%%%%%%%%%%%%
%
\subsection{Physical effects up to $\rmO(T/M)$}
\la{ss:BB}

In a vacuum setting, when heavy quarks are bound inside mesons, their 
momenta are balanced against those of the light constituents, and
thus of order $\Lambdamsbar$ or $m^{ }_{\pi}$. 
In this case, the counting of 
powers of $1/M$ is simple: the explicit terms appearing in the 
denominator represent also the true suppression factors. 
The situation changes at high temperatures $T \gg \Lambdamsbar$, 
because thermal kicks can give the heavy quarks large momenta. 
In fact, in the asymptotic limit of small $\alphas^{ }$, equipartition
asserts that heavy-quark momenta are of order $p^2 \sim M T$. Therefore
we should count covariant derivatives acting on heavy quark fields
as $p \sim\sqrt{MT}$, and heavy quark velocities as shown 
in \eq\nr{Mkin}. 

Given that $p \sim\sqrt{MT} \gg T \gsim \Lambdamsbar $, 
the spatial momenta of 
the heavy quarks should be ``integrated out'', if we want to arrive
at a HQET type formulation for studying non-perturbative effects
from the light parton sector. As a result of the integration, 
we should be left over with HQET type correlation functions, 
multiplied by Wilson coefficients that account for the
effects of the hard modes. Specifically, we may expect
$\langle \vec{v}^{2}_{ } \rangle$ to play the role of 
a (tree-level) Wilson coefficient, which would be 
modified by a multiplicative correction at the loop level.\footnote{%
  The Wilson coefficient generically displays
  a non-vanishing anomalous dimension,
  corresponding to that of the correlation function that it multiplies.
  Its determination entails two matching steps: first, between full QCD
  and the non-relativistic operator in \eq\nr{O1a}; second, between
  thermal NRQCD and thermal HQET from which the momenta
  $p \sim \sqrt{MT}$ have been integrated out. 
  }

For a concrete leading-order implementation of this ideology, we note that
in the static limit, i.e.\ with the action from \eq\nr{S0}, 
the heavy quark propagators reduce to ($\tau > 0$) 
\ba
 \bigl\langle \, 
  \theta^{ }_{\alpha s}(\tau,\vec{x}) \,
  \theta^{*}_{\beta  r}(0,\vec{y})
 \, \bigr\rangle 
  & \stackrel{M\to\infty}{=} & 
  U^{ }_{\alpha\beta}(\tau;0)\,
  \delta^{ }_{sr}\,
  e^{-\tau M} \,  
  \lim_{M\to \infty}
  \int_{\vec{p}} e^{i\vec{p}\cdot(\vec{x}-\vec{y}) 
  - \tau \epsilon^{ }_p}
  % \delta^{(3)}(\vec{x}-\vec{y}\,)
 \;, \la{prop1} \\ 
%%%%%%%%%%%%%%%
 \bigl\langle \, 
  \theta^{ }_{\alpha s}(\beta,\vec{y}) \,
  \theta^{*}_{\beta  r}(\tau,\vec{x})
 \, \bigr\rangle 
  & \stackrel{M\to\infty}{=} &  
  U^{ }_{\alpha\beta}(\beta;\tau)\,
  \delta^{ }_{sr}\, 
  e^{(\tau-\beta) M}
  \lim_{M\to \infty}
  \int_{\vec{q}} e^{i\vec{q}\cdot(\vec{y}-\vec{x}) 
   + (\tau - \beta)\epsilon^{ }_q}
  % \delta^{(3)}(\vec{x}-\vec{y}\,)
 \;, \la{prop2}
\ea
where $U$ is a Wilson line in the fundamental representation
and $\epsilon^{ }_p \equiv p^2/(2M)$ from \eq\nr{S1}. 
Derivatives acting on the propagators yield 
powers of $\vec{p}$ or $\vec{q}$, up to 
radiative corrections from short-distance gauge field fluctuations. 
If we denote by $\vec{k}$ the momentum transfer from the
magnetic or electric field insertion
(so that $\vec{q} = \vec{p} + \vec{k}$)
and shift $\vec{p}$ by $-\vec{k}/2$ for maximal symmetry, 
then the two propagators from \eqs\nr{prop1} and \nr{prop2}
combine into an exponential
\be
 (\tau - \beta)\epsilon^{ }_{\vec{p} + \vec{k}/2}
 - \tau \epsilon^{ }_{\vec{p} - \vec{k}/2}
 = -\beta \Bigl( \epsilon^{ }_p + \frac{k^2}{8 M} \Bigr) 
   +\Bigl(\tau - \frac{\beta}{2}\,\Bigr) \vec{v}\cdot\vec{k}
 \;.   \la{e_sum}
\ee
The part $-\beta\epsilon^{ }_p$ is of $\rmO(1)$ and cannot be 
expanded in, whereas $\beta k^2/ M \sim T/M \ll 1$ represents
a small correction, similar to the Pauli-term in \eq\nr{S1}.  
The last part of \eq\nr{e_sum} generates secular terms 
in the sense discussed
under point (ii) of \se\ref{se:cl},
with $\tau - \beta/2$ representing
an imaginary-time interval and 
$\vec{k} \leftrightarrow -i \nabla$ 
generating spatial translations.

Summarizing these considerations, the term in the exponential that
cannot be expanded is $-\beta\epsilon^{ }_p$.
We normalize the integral over powers of $p^{ }_i$ by the same 
factor appearing in the denominator (i.e.\ $\chi$), 
\be
 \langle p^{ }_i p^{ }_j ... \rangle 
 \; \equiv \; 
 \frac{\lim_{\vec{x}\to\vec{0}}\int_{\vec{p}} p^{ }_i p^{ }_j ... \, 
  e^{ i \vec{p}\cdot\vec{x} - \beta\epsilon^{ }_p }  }{
       \lim_{\vec{x}\to\vec{0}}\int_{\vec{p}} \, 
  e^{ i \vec{p}\cdot\vec{x} - \beta\epsilon^{ }_p }
 }
 \;. \la{pp}
\ee
The Pauli term in \eq\nr{S1} can be expanded in. Furthermore, in 
a perturbative integration out of heavy quark momenta, 
$-\beta k^2/(8 T M) $ in \eq\nr{e_sum} can be expanded in, 
whereas the treatment of the last term in \eq\nr{e_sum} requires 
a decision on how to handle secular terms. 

Let us now apply this recipe to the denominator,  
\ba
 \chi^{ }_{ } & = &
 \int_{\vec{x}} \Bigl\langle \, 
 (\theta^\dagger\theta)(\tau,\vec{x}) \, 
 (\theta^\dagger\theta)(0,\vec{0}) \, 
 e^{-S_1-S_2- ...}
 \, \Bigr\rangle^{ }_0
 \;, 
\ea
where $S^{ }_{1,2}$ are from \eqs\nr{S1} and \nr{S2}, respectively, 
and $\langle ... \rangle^{ }_0$ denotes averaging of the heavy
quark fields with respect to $S^{ }_0$ from \eq\nr{S0}. 
At leading order we find 
\be
 \chi^{ }_{(0)} = 2 e^{-\beta M} \, \int_{\vec{p}} e^{-\beta \epsilon^{ }_p} \,
 \tr \bigl\langle U(\beta;0) \rangle
 \;. \la{chi0}
\ee
There is no first-order correction, because the Pauli term
from \eq\nr{S2} vanishes by the spin trace. There are corrections
at $1/M^2$, but these are not needed, as we want to extract 
effects of $\rmO(T/M)$. In short, 
$\chi$ can be replaced by $\chi^{ }_{(0)}$ at $\rmO(T/M)$.
The same arguments apply also in the numerator: 
at $\rmO(T/M)$, we do not need to worry about
the expansion of the Pauli term in \eq\nr{S1} nor of \eq\nr{S2}. 

With this recipe, the 
correlator $G^{ }_{\{01a\}}$ from \eq\nr{G01a} 
of naive $\rmO(1/M)$ vanishes at leading order,
because it contains one spatial derivative and is thus 
proportional to $\langle \vec{p} \rangle = \vec{0}$
or $\langle \vec{q} \rangle = \vec{0}$. However, 
if we expand the last term of \eq\nr{e_sum} into the correlator, 
we get a contribution proportional to 
$\langle \vec{v}^{2}_{ }\rangle \sim T/M$ which does not vanish. 
This contribution contains the prefactor $\tau - \beta/2$ and
is a secular term in the sense discussed 
under point (ii) of \se\ref{se:cl}.
As explained there, we do not think that 
it is physically sensible to consider these effects, 
if the goal is to match onto a Langevin description. 

The correlator $G^{ }_{\{02e\}}$ from \eq\nr{G02a} does not vanish, 
but evaluates at $\rmO(T/M)$ to 
\be
 G^{ }_{\{02e\}}(\tau) = 
 \frac{5 
 \bigl\langle \vec{v}^{2}_{ } \bigr\rangle
  }{3 }
 \frac{ 
 g^2 \,\tr \bigl\langle
 U(\beta;\tau) E^{ }_i(\tau) 
 U(\tau;0) E^{ }_i(0) 
 \bigr\rangle
 }
 { 
 3\, \tr \bigl\langle U(\beta;0) \rangle
 }
 \;. \la{G_02a} 
\ee
We have suppressed spatial coordinates, as they are all the same. 
This is just $-5 \bigl\langle \vec{v}^{2}_{ } \bigr\rangle / 3$
times the leading-order correlator originating from 
\eq\nr{G_00}, and corresponds to the effects discussed
below \eq\nr{acc2}. As elaborated upon there, the corresponding
effects can be eliminated by going over to covariant momenta
in the Langevin description. 

Finally, there is 
genuine effect of $\rmO(T/M)$ from $G^{ }_{1a1a}$ in \eq\nr{G1a1a}, 
\be
 G^{ }_{1a1a}(\tau) = 
 \frac{2  \,\bigl\langle \vec{v}^{2}_{ } \bigr\rangle
 }{3 }
 \frac{ 
 g^2 \,\tr \bigl\langle
 U(\beta;\tau) B^{ }_i(\tau) 
 U(\tau;0) B^{ }_i(0) 
 \bigr\rangle
 }
 { 
 3\, \tr \bigl\langle U(\beta;0) \rangle
 }
 \;. \la{G_1a1a}
\ee
Both expectation values are real in the physical ground state, 
so we may add a real part in order to eliminate noise, and 
define somewhat more explicitly   
\be
 G^{ }_{\!\B}(\tau) \; \equiv \;
 \frac{
        \sum_i \re\tr\langle U(\beta;\tau)\, gB^{ }_i(\tau)\, 
                      U(\tau;0) \, gB^{ }_i(0)\, \rangle
      }{
        3 \re\tr\langle U(\beta;0) \rangle
      }
 \;. \la{GB_def}
\ee
Then the full transport coefficient reads 
\be
 \kappa^{ }_\rmi{tot} \;\simeq\; 
 \kappa^{ }_{\!\E} + 
 \frac{2}{3} \langle \vec{v}^{2}_{ } \rangle \, 
 \kappa^{ }_{\!\B}
 \;, \la{kappa_full}
\ee
where $ \langle \vec{v}^{2}_{ } \rangle $ 
is given by \eq\nr{Mkin} and the discussion below it, and 
$\kappa^{ }_{\!\E,\B}$ are the transport coefficients corresponding
to $G^{ }_{\!\E,\B}$, respectively. 

%%%%%%%%%%%%%%%%%%%%%%%%%%%%% SECTION %%%%%%%%%%%%%%%%%%%%%%%%%%%%%%%%%%%%
%
\section{Perturbative evaluation}
\la{se:pert} 

The purpose of this section is to look in more detail
into the correlator $G^{ }_{\!\B}$ defined in \eq\nr{GB_def}, determining
both the temporal correlator and the transport coefficient at 
leading order. 

%%%%%%%%%%%%%%%%%%%%%%%%%%% SUBSECTION %%%%%%%%%%%%%%%%%%%%%%%%%%%%%%%%%%%
%
\subsection{Temporal correlator}

Let us start by recalling the form of the electric correlator
from \eq\nr{G_00}, 
\ba
%%%%%%%%%%
 \,\tr \bigl\langle
 U(\beta;\tau) E^{ }_i(\tau) 
 U(\tau;0) E^{ }_i(0) 
 \bigr\rangle
 & \stackrel{\rmO(g^0)}{=} & 
 \frac{\Nc^2 - 1}{2}
 \Tint{K}
 \frac{  
 ( 3 \omega_n^2 + \vec{k}^{2}_{ } )
  e^{i \omega_n \tau}  }{\omega_n^2 + \vec{k}^{2}_{ } }
%%%%%%%%%%%
 \nn 
 & \stackrel{\omega_n^2 = \omega_n^2 + \vec{k}^{2}_{ } - \vec{k}^{2}_{ }}{=} & 
 - 
 (\Nc^2 - 1)
 \Tint{K}
 \frac{  
 \vec{k}^{2}_{ }
 \, e^{i \omega_n \tau}  }{\omega_n^2 + \vec{k}^{2}_{ } }
 \;, 
\ea
where in the second step a contact term vanishing 
in dimensional regularization was omitted. 
For $G^{ }_{\!\B}$ the corresponding evaluation gives
\ba
%%%%%%%%%%
 \,\tr \bigl\langle
 U(\beta;\tau) B^{ }_i(\tau) 
 U(\tau;0) B^{ }_i(0) 
 \bigr\rangle
 & \stackrel{\rmO(g^0)}{=} & 
 \frac{\Nc^2 - 1}{2}
 \Tint{K}
 \frac{  
 \epsilon^{ }_{ijl}\epsilon^{ }_{imn}\,
 k^{ }_j k^{ }_m\, 
 \delta^{} _{ln} e^{i \omega_n \tau}  }{\omega_n^2 + \vec{k}^{2}_{ } }
%%%%%%%%%%%
 \nn 
 & = & 
 (\Nc^2 - 1)
 \Tint{K}
 \frac{  
 \vec{k}^{2}_{ }
 \, e^{i \omega_n \tau}  }{\omega_n^2 + \vec{k}^{2}_{ } }
 \;.
\ea
Recalling the overall minus sign 
in the colour-electric correlator, cf.\ \eq\nr{G_00}, 
the correlators $G^{ }_{\!\E}$ and $G^{ }_{\!\B}$ agree at leading order. 
Adding the denominator, 
carrying out the Matsubara sum, 
and denoting by $G^{(n)}_{\B}$ the
contribution of order $g^n_{ }$, we get~\cite{kappaE} 
\ba
 G^{(2)}_{\!\B}(\tau)
 & = & 
 \frac{g^2\CF^{ }}{3}
 \int_\vec{k} 
 \bigl[ e^{\tau k} + e^{(\beta - \tau) k} \bigr] \, k \,  \nB^{ }(k)  
 \nn 
 & = &
 g^2 \CF^{ }\, \pi^2 T^4 
 \biggl[
 \frac{\cos^2(\pi \tau T)}{\sin^4(\pi \tau T)}
 +\frac{1}{3\sin^2(\pi \tau T)} 
 \biggr]
 \;, 
\ea
where $\nB^{ }$ is the Bose distribution. 
This can serve as a normalization for 
lattice results. 

%%%%%%%%%%%%%%%%%%%%%%%%%%% SUBSECTION %%%%%%%%%%%%%%%%%%%%%%%%%%%%%%%%%%%
%
\subsection{Transport coefficient}

The determination of the transport coefficient 
$\kappa^{ }_{\!\B}$
is non-trivial so we 
give some details, following the presentation 
for $G^{ }_{\!\E}$ and $\kappa^{ }_{\!\E}$ in ref.~\cite{rhoE}. 

%%%%%%%%%%%%%%%%%%%%%%%%%%% SUBSUBSECTION %%%%%%%%%%%%%%%%%%%%%%%%%%%%%%%%
%
\subsubsection{Setup}

Formally, after the Fourier transformation 
\be
 \tilde G^{ }_{\!\B}(\omega^{ }_n) = 
 \int_0^\beta \! {\rm d}\tau \, e^{i \omega_n \tau} \, G^{ }_{\!\B}(\tau)
 \;, \la{tildeGB_def}
\ee
we may extract the spectral function
\be
 \rho^{ }_{\!\B}(\omega) = 
 \im \tilde G^{ }_{\!\B}(\omega^{ }_n \to -i [\omega + i 0^+_{ }]\,)
 \;. \la{rhoB_def}
\ee
The transport coefficient is given by 
\be
 \kappa^{ }_{\!\B} = \lim_{\omega\to 0} 
 \frac{2T\rho^{ }_{\!\B}(\omega)}{\omega}
 \;, \la{kappaB_def}
\ee
where the factor $2T/\omega$ transforms a spectral function
(commutator) into the anticommutator appearing 
in \eq\nr{kappa_M_w}.

As is normally the case with transport coefficients, 
practical computations necessitate 
a resummation of the perturbative series. We may write
\be
 \kappa^{ }_{\!\B} = \kappa^{\rmii{QCD,expanded}}_{\!\B}
               - \kappa^{\rmii{HTL,expanded}}_{\!\B}
               + \kappa^{\rmii{HTL,full}}_{\!\B}
 \;. \la{resum}
\ee
Here ``expanded'' denotes an unresummed computation,
i.e.\ a naive expansion in the coupling~$g^2$; 
HTL stands for Hard Thermal Loop resummation~\cite{ht1,ht2,ht3,ht4}; 
and the subtraction removes the danger of double counting. 
If there were no IR divergences, or if we could compute
to all orders in the expansion, 
the latter two terms would cancel against each other. 

The full spectral function, 
computed as in ref.~\cite{rhoE}, 
contains a vacuum part proportional to $\omega^3$ as well as 
a thermal part with a complicated functional dependence on $\omega$.
According to \eq\nr{kappaB_def}, we only need 
the part linear in $\omega$ at small $\omega$, which can only
originate from thermal corrections. Therefore, in the following,
we omit those vacuum corrections which 
are of NLO in the weak-coupling expansion
(for completeness we do display the LO vacuum term
in \eq\nr{rhoB_lo}).

We carry out the computation with 
dimensional regularization, in $D\equiv 4 - 2\epsilon$ dimensions.  
Introducing a scale parameter $\mu$, the $\msbar$ scale is defined as 
$
 \bmu^2 \equiv 4\pi \mu^2 e^{-\gammaE} 
$.
The Levi-Civita-symbol is written as 
$
 \epsilon^{ }_{ijk}\epsilon^{ }_{lmn}
 \; \equiv \; 
 \delta^{ }_{il}\delta^{ }_{jm}\delta^{ }_{kn}
 + \delta^{ }_{im}\delta^{ }_{jn}\delta^{ }_{kl}
 + \delta^{ }_{in}\delta^{ }_{jl}\delta^{ }_{km}
 - \delta^{ }_{il}\delta^{ }_{jn}\delta^{ }_{km}
 - \delta^{ }_{im}\delta^{ }_{jl}\delta^{ }_{kn}
 - \delta^{ }_{in}\delta^{ }_{jm}\delta^{ }_{kl}
$.
Often we are faced with 
$
 \delta^{ }_{il}
 \delta^{ }_{jm}
 \epsilon^{ }_{ijk}\epsilon^{ }_{lmn}
 = 
 (D-3)(D-2)\delta^{ }_{kn}
$.
Following ref.~\cite{rhoE}, we denote a massless propagator as 
\be
 G(x^{ }_0,\vec{x})
 \; \equiv \; 
 \Tint{K} \frac{e^{i (k^{ }_n x^{ }_0 + \vec{k}\cdot\vec{x})}}{K^2}
 \;. 
\ee

%%%%%%%%%%%%%%%%%%%%%%%%%%% SUBSUBSECTION %%%%%%%%%%%%%%%%%%%%%%%%%%%%%%%%
%
\subsubsection{QCD contributions}

%%%%%%%%%%%%%%%%%%%%%%%%% FIGURE %%%%%%%%%%%%%%%%%%%%%%%%%%%%%%%%%%%%%%%%%
%
\begin{figure}[t]
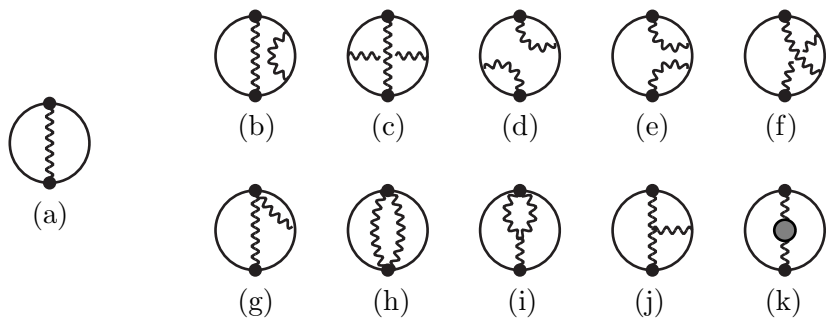


\hspace*{1.5cm}%
\begin{minipage}[c]{3cm}
\begin{eqnarray*}
&& 
 \hspace*{-1cm}
 \EleA 
\\[1mm] 
&& 
 \hspace*{-0.6cm}
 \mbox{(a)} 
\end{eqnarray*}
\end{minipage}%
\begin{minipage}[c]{10cm}
\begin{eqnarray*}
&& 
 \hspace*{-1cm}
 \EleB \quad\; 
 \EleBB \quad\; 
 \EleC \quad\; 
 \EleD \quad\; 
 \EleE \quad\; 
\\[1mm] 
&& 
 \hspace*{-0.6cm}
 \mbox{(b)} \hspace*{1.26cm}
 \mbox{(c)} \hspace*{1.26cm}
 \mbox{(d)} \hspace*{1.26cm}
 \mbox{(e)} \hspace*{1.26cm}
 \mbox{(f)} 
\\[5mm] 
&& 
 \hspace*{-1cm}
 \EleF \quad\; 
 \EleG \quad\; 
 \EleH \quad\; 
 \EleI \quad\; 
 \EleJ \quad 
\\[1mm] 
&& 
 \hspace*{-0.6cm}
 \mbox{(g)} \hspace*{1.3cm}
 \mbox{(h)} \hspace*{1.28cm}
 \mbox{(i)} \hspace*{1.28cm}
 \mbox{(j)} \hspace*{1.32cm}
 \mbox{(k)} 
\end{eqnarray*}
\end{minipage}

\caption[a]{\small 
The LO and NLO graphs contributing to the colour-magnetic correlator, 
$G^{ }_{\!\B}$, defined in \eq\nr{GB_def}. 
The big circle denotes a Polyakov loop; 
the small dots colour-magnetic field strengths; 
and the grey blob the 1-loop gauge field self-energy.} 
\la{fig:GB}
\end{figure}
%
%%%%%%%%%%%%%%%%%%%%%%%%%%%%%%%%%%%%%%%%%%%%%%%%%%%%%%%%%%%%%%%%%%%%%%%%%%

We start by extracting the part denoted by 
$ \kappa^{\rmii{QCD,expanded}}_{\!\B} $ 
in \eq\nr{resum}. 
This originates from graphs that are of NLO, i.e.\ $\rmO(g^4)$, 
with diagrams as shown in \fig\ref{fig:GB}. 
We have carried out the computation in a general covariant gauge, 
verifying its gauge independence; below, graph-by-graph results
are listed for the Feynman gauge. 

The LO graph gives 
\be
 \delta^{ }_\rmi{(a)} G^{ }_{\!\B}(\tau)  
 = \frac{g^2 \CF^{ }(D-3)(D-2)}{3}
  (-\nabla^2) G(\tau,\vec{0})
 \;.  \la{GB_LO}
\ee
This leads to 
\be
 \delta^{ }_\rmi{(a)} \rho^{ }_{\!\B}(\omega)
 \; \stackrel{D=4}{=} \;
 \frac{g^2 \CF^{ } \omega^3}{6\pi}
 \;, \la{rhoB_lo}
\ee
so there is no contribution to $\kappa^{ }_{\!\B}$.
We also need the NLO correction to the denominator, 
\be
 \re\tr\langle U(\beta;0) \rangle = 
 \Nc^{ }\, 
 \bigl\{ 
  1 - g^2 \CF^{ }\int_0^\beta \! {\rm d}\tau' 
                 \int_0^{\tau'} \! {\rm d}\tau'' \, 
                 G(\tau'-\tau'',\vec{0})
 \bigr\}
 \;. \la{denom}
\ee

Consider next the diagrams (b-f). In Feynman gauge, the diagrams (d,e,f)
are absent. The diagrams (b,c) yield a term proportional to $\CF^2$, which
cancels when the LO term in \eq\nr{GB_LO} is combined with the 
NLO correction from \eq\nr{denom}. The combined effect is thus 
proportional to $\Nc^{ }\CF^{ }$, 
\be
 \delta^{ }_\rmi{(a)}
 G^{ }_{\!\B}(\tau)
 \times 
 \frac{\chi^{ }_\rmiii{LO}}{\chi^{ }_\rmiii{NLO}}
 + 
 \delta^{ }_\rmi{(b-f)}
 G^{ }_{\!\B}(\tau)
 \; = \;
 \frac{g^4 \Nc^{ } \CF^{ }(D-3)(D-2)}{3} \, \I^{ }_4(\tau)
 + \rmO(g^6) 
 \;, 
\ee
where 
\be
 \I^{ }_4(\tau) \; \equiv \;  
 \Tint{K} \frac{k^2 e^{i k_n\tau}}{K^2}
 \Tint{Q'} \frac{e^{i q_n\tau} - 1}{q_n^2 Q^2} 
\ee
was defined in \eq(3.28) of ref.~\cite{rhoE}. 

Proceeding further, graph (g) is absent in Feynman gauge. 
Graphs (h) and (i) give
\ba
 \delta^{ }_\rmi{(h)}
 G^{ }_{\!\B}(\tau) & = & 
 \frac{g^4 \Nc^{ } \CF^{ }(D-3)(D-2)(D-1)}{6} \, \I^{ }_1(\tau)
 \;, \\ 
 \delta^{ }_\rmi{(i)}
 G^{ }_{\!\B}(\tau) & = & - 
 g^4 \Nc^{ } \CF^{ }(D-3)(D-2) \, \I^{ }_2(\tau)
 \;, 
\ea
where $\I^{ }_{1,2}$ were 
defined in \eqs(3.25,3.26) of ref.~\cite{rhoE}, 
\be
 \I^{ }_1(\tau) \; \equiv \; 
 \Tint{K,Q}
 \frac{e^{i k_n\tau}}{Q^2(K-Q)^2}
 \;, \quad
%%%%
 \I^{ }_2(\tau) \; \equiv \; 
 \Tint{K,Q} 
 \frac{k^2 e^{i k_n\tau}}{K^2 Q^2(K-Q)^2}
 \;. \la{I1I2}
\ee
Graph~(j) is a fairly complicated one, yielding
\ba
 \delta^{ }_\rmi{(j)}
 G^{ }_{\!\B}(\tau) & = & 
 \frac{g^4 \Nc^{ } \CF^{ }(D-3)(D-2)}{6} \, \I^{ }_6(\tau)
 \;, \\ 
 \I^{ }_6(\tau) & \equiv & 
 \biggl[\int_{\tau}^{\beta} \! {\rm d}\tau'  - 
 \int_{0}^{\tau} \! {\rm d}\tau' \biggr] 
 \int_X
   G(x^{ }_0-\tau',\vec{x}) 
  \nn 
  & \times &  
   \Bigl[ 
          \partial^{ }_i G(x^{ }_0-\tau,\vec{x}) 
          \,\partial^{ }_0 \partial^{ }_i  G(x^{ }_0,\vec{x}) 
                - 
          \partial^{ }_0 \partial^{ }_i G(x^{ }_0-\tau,\vec{x}) 
          \,\partial^{ }_i G(x^{ }_0,\vec{x}) 
   \Bigr]
 \;, \hspace*{4mm} \la{I6_def}
\ea
where $X \equiv (x^{ }_0,\vec{x})$.
The function $\I^{ }_6$ is similar but not identical 
to $\I^{ }_5$, defined in \eq(3.18) of ref.~\cite{rhoE}; 
in fact we get $\I^{ }_6$ if we set $D\to 3$ in a prefactor
appearing in $\I^{ }_5$, 
and carry out a partial integration. 
Finally, the self-energy graph~(k) yields 
\ba
 \delta^{ }_\rmi{(k)}
 G^{ }_{\!\B}(\tau) & = & 
 \frac{g^4 \CF^{ }(D-3)}{3} \, \biggl\{ 
  (D-2)\Nc^{ }
  \biggl[
  - \frac{(D-2)(D-1)}{2}\, \I^{ }_0(\tau)
  + 2\, \I^{ }_2(\tau) 
  + 2\, \I^{ }_7(\tau)
  \biggr]
%%%%%%%%%%%%
 \nn 
 & & 
 + \, \Nf^{ }
  \biggl[
     (D-2)(D-1)\, \I^{ }_{\{0\}}(\tau) 
   - (D-2)\, \I^{ }_{\{2\}}(\tau)
   - 4\, \I^{ }_{\{7\}}(\tau)
  \biggr]
  \biggr\}
 \;, 
\ea
where the four-momentum $Q$ is fermionic in 
$\I^{ }_{\{0\}}$, 
$\I^{ }_{\{2\}}$ and
$\I^{ }_{\{7\}}$; 
we have made use of the identity
\be
 \int_{\vec{k}} \frac{k^2}{K^4} = \frac{D-1}{2}
 \int_{\vec{k}} \frac{1}{K^2}
\ee
in order to express results in terms of the factorizable structure
$\I^{ }_{0}$ 
defined as
\be
 \I^{ }_0(\tau) \; \equiv \; 
 \Tint{K,Q}
 \frac{e^{i k_n\tau}}{K^2(K-Q)^2} 
 \;, 
\ee 
and we have denoted 
\be
 \I^{ }_7(\tau) \; \equiv \; - \lim_{\lambda\to 0}
 \frac{{\rm d}}{{\rm d}\lambda^2}
 \Tint{K,Q} 
 \frac{[k^2 q^2 - (\vec{k}\cdot\vec{q}\,)^2 ] e^{i k_n\tau}}
      {(K^2+\lambda^2)Q^2(K-Q)^2} 
 \;. \la{I7_def}
\ee
Even if not obvious at first sight, the thermal part of 
$\I^{ }_7$ turns out to be closely related to the thermal part of 
$\I^{ }_3$, defined in \eq(3.27) of ref.~\cite{rhoE}. 

In order to extract $\kappa^{ }_{\!\B}$, we denote, 
in accordance with \eqs\nr{tildeGB_def}--\nr{kappaB_def},  
\be
 \tilde\I^{ }_i(\omega) 
 \; \equiv \; 
 \im 
  \int_0^\beta \! {\rm d}\tau \, e^{i \omega_n \tau} \, \I^{ }_i(\tau) 
  |^{ }_{\omega^{ }_n \to -i [\omega + i 0^+_{ }] }
 \;, \quad
 \kappa^{ }_i
 \; \equiv \; 
 \lim_{\omega\to 0} \frac{2T\,\tilde\I^{ }_i(\omega)}{\omega}
 \;. 
\ee
The functions $\tilde I^{ }_{1,2,4}(\omega)$ are given 
in \eqs(A.54,55,57), respectively, of ref.~\cite{rhoE}, 
whereas $\tilde I^{ }_{0}$ can be extracted from \eq(B.9), 
multiplied by the tadpole 
$\Tinti{Q}\frac{1}{Q^2} = \frac{T^2}{12}$
or
$\Tinti{\{Q\}}\frac{1}{Q^2} = -\frac{T^2}{24}$.
This leads to 
\be
 \kappa^{ }_0 = 
 \frac{T^3}{24\pi}
 \;, \quad
 \kappa^{ }_1 = 
 \kappa^{ }_2 = 
 \frac{T^3}{12\pi}
 \;, \quad
 \kappa^{ }_4 = - \frac{T^3}{12\pi}
 \;, \quad
 \kappa^{ }_{\{0\}} =
 - \frac{T^3}{48\pi}
 \;, \quad
 \kappa^{ }_{\{2\}} =
 - \frac{T^3}{24\pi}
 \;.
\ee
For the function $\tilde\I^{ }_6$, originating from \eq\nr{I6_def},  
we find the thermal part
\ba
 \tilde\I^{ }_6(\omega) & \supset & 
 \frac{1}{8\pi^3}
 \int_0^\infty \! {\rm d}q \, \nB^{ }(q) \, \mathbbm{P} \biggl\{ 
    4 q \omega \biggl( 1 + \frac{\omega^2}{\omega^2 - q^2} \biggr) 
  + 5 \omega^2 \ln \biggl| \frac{q+\omega}{q-\omega} \biggr|
  \nn 
 & + & 
   \frac{2\omega^4}{q}
   \biggl[
    \frac{1}{q+\omega} \ln \frac{q+\omega}{\omega} -  
    \frac{1}{q-\omega} \ln \frac{|q-\omega|}{\omega}
   \biggr]
 \biggr\} 
 \;, 
\ea
where $\mathbbm{P}$ denotes a principal value. 
The small-$\omega$ limit amounts to 
\be
 \kappa^{ }_6 = \frac{T^3}{6\pi}
 \;. 
\ee
The function $\tilde\I^{ }_7(\omega)$, 
originating from \eq\nr{I7_def}, can be obtained in close analogy
with the discussion in appendices A.1-3 of ref.~\cite{rhoE}, and its 
thermal part agrees, up to a sign, with that of $\tilde\I^{ }_3(\omega)$
in \eq(A.56) of ref.~\cite{rhoE}: 
\be
 \tilde\I^{ }_7(\omega) 
 \supset
 \frac{1}{16\pi^3}
 \int_0^{\infty} \! {\rm d}q \, \nB^{ }(q) \, 
 \biggl\{ 
   q^2 \ln\biggl| \frac{q + \omega}{q - \omega} \biggr|
  + q \omega \ln \biggl| \frac{q^2 - \omega^2}{\omega^2} \biggr|
 \biggr\} 
 \;. 
\ee
In the fermionic $\tilde\I^{ }_{\{7\}}$, we need to replace
$\nB^{ }\to -\nF^{ }$. The corresponding transport coefficients 
amount to 
\ba 
 \kappa^{ }_7 & \approx & \frac{T^3}{24\pi}
 \biggl[
  \ln\biggl(\frac{T}{\omega}\biggr) 
  + 2 - \gammaE + \frac{\zeta'(2)}{\zeta(2)} 
 \biggr]
 \;, \\ 
 \kappa^{ }_{\{7\}} & \approx & - \frac{T^3}{48\pi}
 \biggl[
  \ln\biggl(\frac{2T}{\omega}\biggr)
  + 2 - \gammaE + \frac{\zeta'(2)}{\zeta(2)} 
 \biggr]
 \;, 
\ea
where ``$\approx$'' indicates that the limit $\omega\to 0$ has not 
been taken inside the logarithm. 

Summing up the effects from all the diagrams, we obtain
\ba
 \kappa_{\!\B}^\rmii{QCD,expanded} & \approx  & 
 \frac{g^4 \CF^{ } T^3}{18\pi}
 \biggl\{ 
  \Nc^{ }\, 
            \biggl[
  \ln\biggl(\frac{T}{\omega}\biggr) 
  + 1 - \gammaE + \frac{\zeta'(2)}{\zeta(2)} 
            \biggr]
%%%%%%%
 \nn 
 &  &
 \hspace*{1.2cm} + \, 
  \frac{\Nf^{ }}{2}
            \biggl[ 
  \ln\biggl(\frac{2T}{\omega}\biggr)
  + \frac{3}{2} - \gammaE + \frac{\zeta'(2)}{\zeta(2)} 
            \biggr]
 \biggr\} 
 \;. \la{kappaB_QCD}
\ea
Clearly the result is IR-divergent for $\omega\to 0$, 
and necessitates resummation. 

%%%%%%%%%%%%%%%%%%%%%%%%%%% SUBSUBSECTION %%%%%%%%%%%%%%%%%%%%%%%%%%%%%%%%
%
\subsubsection{Hard Thermal Loop resummation}

Plasma effects from the colour-electric scale, 
$
 \mE^2 
 \stackrel{\rmii{$D=4$}}{=} g^2 T^2 \bigl( \frac{\Nc}{3} + \frac{\Nf}{6} \bigr)
$, 
can be incorporated by carrying out HTL resummation~\cite{ht1,ht2,ht3,ht4}. 
The ingredients needed for the current problem are 
described in appendix~B of ref.~\cite{rhoE}, and can be 
adapted to $G^{ }_{\!\B}$ with minor modifications. 
For the ``expanded'' version we obtain, 
in analogy with \eq(B.14) of ref.~\cite{rhoE}, 
\ba
 \kappa^{\rmii{HTL,expanded}}_{\!\B} & \approx & 
 \frac{2 g^2 \CF^{ } \mE^2 T (D-3)}{3}
 \frac{\mu^{-2\epsilon}}{8\pi}
 \biggl(
  \frac{1}{\epsilon} + \ln\frac{\bmu^2}{4\omega^2} - 1  
 \biggr)
 \;. \la{kappaB_HTL_expanded}
\ea
The ``full'' HTL computation does not regularize $G^{ }_{\!\B}$, 
as colour-magnetic fields play a role. We introduce an intermediate
{\it ad hoc} IR regulator, $\mG^{ }$, as a colour-magnetic scale, 
and represent the full HTL result, 
in analogy with \eq(B.18) of ref.~\cite{rhoE}, as
\ba
 \kappa^{\rmii{HTL,full}}_{\!\B} & = & 
 \frac{2 g^2 \CF^{ } \mE^2 T (D-3)}{3}
 \lim_{\mG^2\to 0}
 \int_{\vec{k}} 
 \frac{\vec{k}^2 \pi \delta(\vec{k}\cdot\vec{v})}{(\vec{k}^2 + \mG^2)^2}
 \nn 
 & = & 
 \frac{2 g^2 \CF^{ } \mE^2 T (D-3)}{3}
 \lim_{\mG^2\to 0}
 \frac{\mu^{-2\epsilon}}{8\pi}
 \biggl(
  \frac{1}{\epsilon} + \ln\frac{\bmu^2}{\mG^2} - 1  
 \biggr)
 \;. \la{kappaB_HTL_full}
\ea
Clearly the result is IR divergent; it is rendered finite by 
non-perturbative dynamics at the colour-magnetic scale 
$\sim \alpha^{ }_s\Nc T$~\cite{linde}. 
The difference 
$
 \kappa^{\rmii{HTL,full}}_{\!\B}
 - 
 \kappa^{\rmii{HTL,expanded}}_{\!\B}
$, 
appearing in \eq\nr{resum}, 
is however UV finite as it should (i.e.\ $1/\epsilon$ cancels), 
and effectively replaces the $\ln\omega$'s in 
\eq\nr{kappaB_QCD} with logarithms of the colour-magnetic scale. 

%%%%%%%%%%%%%%%%%%%%%%%%%%% SUBSUBSECTION %%%%%%%%%%%%%%%%%%%%%%%%%%%%%%%%
%
\subsubsection{Summary}

Summing up the effects from \eqs\nr{kappaB_QCD}, 
\nr{kappaB_HTL_expanded}, \nr{kappaB_HTL_full}
according to \eq\nr{resum}, and replacing the artificial 
scale $\mG^{ }$ by the physical colour-magnetic scale 
$\alphas\Nc T$, which however is associated with 
a process-dependent non-perturbative constant, $c^{ }_\B$, 
we obtain
\ba
 \kappa_{\!\B}^\rmii{ } & =  & 
 \frac{g^4 \CF^{ } T^3}{18\pi}
 \biggl\{ 
  \Nc^{ }\, 
            \biggl[
  \ln\biggl(\frac{2 c^{ }_\B }{\alpha_s\Nc}\biggr) 
  + 1 - \gammaE + \frac{\zeta'(2)}{\zeta(2)} 
            \biggr]
%%%%%%%
 \nn 
 &  &
 \hspace*{1.2cm} + \, 
  \frac{\Nf^{ }}{2}
            \biggl[ 
  \ln\biggl(\frac{4 c^{ }_\B }{\alpha_s\Nc}\biggr)
  + \frac{3}{2} - \gammaE + \frac{\zeta'(2)}{\zeta(2)} 
            \biggr]
 \biggr\} 
 \; + \; \rmO(g^5) 
 \;. \la{kappaB_full}
\ea
Even if $c^{ }_\B$ remains unknown, 
the coefficients of the logarithms are unambiguously predicted, 
as they characterize contributions from a ratio of scales. 
It is worth remarking that 
$
 1 -\gammaE + \frac{\zeta'(2)}{\zeta(2)} \approx -0.147
$, so the argument of the first logarithm should exceed 1.16
for a sensible result. 

The expression for $\kappa^{ }_{\!\B}$ is quite similar
to that for $\kappa^{ }_{\!\E}$~\cite{kappa_lo}, with the difference
that in the latter IR sensitivity is regularized by 
$\mE$ rather than $\alphas\Nc^{ }T$. In the extreme weak-coupling
limit, $\alphas\Nc^{ }T \ll \mE$, so we might expect 
$\kappa^{ }_{\!\B}$ to be larger than $\kappa^{ }_{\!\E}$, however 
the difference is inside a logarithm so this argument is weak.  
In general, colour-magnetic corrections tend to be 
large compared with those from the scale $\mE$ 
(cf.,\ e.g.,\ ref.~\cite{masses} for a review), 
however NLO corrections to $\kappa^{ }_{\!\E}$ 
are large and positive~\cite{kappa_nlo}, so it is hard to 
anticipate whether there is a clear hierarchy 
between $\kappa^{ }_{\!\E}$ and $\kappa^{ }_{\!\B}$
at practically reachable temperatures. 

% \vspace*{-2mm}

%%%%%%%%%%%%%%%%%%%%%%%%%%%%% SECTION %%%%%%%%%%%%%%%%%%%%%%%%%%%%%%%%%%%%
%
\section{Conclusions and outlook}
\la{se:concl} 

The purpose of the present paper has been to consider effects
of relative order 
$\rmO(T/M)$ to the heavy quark momentum diffusion coefficient, 
$\kappa$. 
We have argued that one correction originates from 
a colour-magnetic correlator dressing a Polyakov loop
(cf.\ \eq\nr{GB_def}), and that the corresponding contribution
to $\kappa$, denoted by $\kappa^{ }_{\!\B}$ (cf.\ \eq\nr{kappa_full}),  
is non-perturbative already at leading order of 
the weak-coupling expansion (cf.\ \eq\nr{kappaB_full}).
Apart from the known dramatic increase of 
$\kappa^{ }_{\!\E}$ 
through interactions~\cite{kappa_nlo,lat2,lat25,lat3,lat4,lat5,lat6}, 
additional $\rmO(T/M)$-effects originating from~$\kappa^{ }_{\!\B}$
could help to explain why charm quarks show fast kinetic
equilibration at temperatures not much above the confinement scale
(cf.,\ e.g.,\ ref.~\cite{pheno}). 

It should be stressed, however,  
that $\kappa^{ }_{\!\B}$ does not represent
the only potential $\rmO(T/M)$ effect.
On the side of rate coefficients, 
further corrections of $\rmO(T/M)$ could be eliminated
by taking care to avoid secular terms 
(cf.\ point (ii) in \se\ref{se:cl}), 
by using covariant momenta in Langevin simulations 
(cf.\ point (iii) in \se\ref{se:cl}), 
and by including corrections in the relation of $\kappa$ and $\eta$
(cf.\ point (iv) in \se\ref{se:cl}).
On the side of dispersive effects,  
a correction originates from 
the difference between the vacuum pole mass $M$ and the thermal kinetic mass
$M^{ }_\rmi{kin}$, even if it is difficult to quantify this effect, 
given that $M$ is  
ambiguous by $\sim\rmO(\Lambdamsbar)$; in practice, 
$M^{ }_\rmi{kin}$ should probably be treated as a fit parameter.  
In spite of these additional ingredients, 
we hope that an estimate of $\kappa^{ }_{\!\B}$ 
could give a fair impression about the size of finite-mass corrections
to heavy quark rate observables. 

As far as a lattice study of $\kappa^{ }_{\!\B}$ goes, 
the prospects look quite good. Previous investigations of the 
colour-electric correlator~\cite{lat2,lat25,lat3,lat4,lat5,lat6}
show that a signal can be obtained, and a continuum
extrapolation is feasible. 
On the aspect of renormalization, where 
only perturbative factors have been worked out 
for $\kappa^{ }_{\!\E}$~\cite{renormE}, the non-perturbative 
level has been reached for a particular discretization 
of colour-magnetic fields~\cite{renormB}. Conceivably, 
gradient flow~\cite{wilson} could offer further tools for studying
the colour-magnetic correlator, along the lines 
discussed in ref.~\cite{renormflow}, 
and physical insight could be obtained from classical lattice
gauge theory simulations~\cite{clgt,noneq}. 
Finally, it might be worth considering
whether the colour-magnetic correlator captures some dispersive
effects as well, 
in analogy with the colour-electric one~\cite{shift}.

% \vspace*{-2mm}

%%%%%%%%%%%%%%%%%%%%%%%%%%% SECTION %%%%%%%%%%%%%%%%%%%%%%%%%%%%%%%%%%
%
\section*{Acknowledgements}

M.L.\ thanks Debasish Banerjee for suggesting the topic many years ago
and for his continued interest in it, 
and Saumen Datta for helpful discussions. 
Our work was partly supported by the Swiss National Science Foundation
(SNF) under grant 200020B-188712.

\small{
%%%%%%%%%%%%%%%%%%%%%%%%%% BIBLIO %%%%%%%%%%%%%%%%%%%%%%%%%%%%%%%%%%%%%%%%%
%

}

\end{document}